\documentclass[12pt]{article}
	
	\newcommand{\blind}{0}
	
	\usepackage[margin=1in]{geometry}

    \makeatletter
    \renewcommand\section{\@startsection {section}{1}{\z@}%
                                       {-3.5ex \@plus -1ex \@minus -.2ex}%
                                       {2.3ex \@plus.2ex}%
                                       {\normalfont\fontfamily{phv}\fontsize{16}{19}\bfseries}}
    \renewcommand\subsection{\@startsection{subsection}{2}{\z@}%
                                         {-3.25ex\@plus -1ex \@minus -.2ex}%
                                         {1.5ex \@plus .2ex}%
                                         {\normalfont\fontfamily{phv}\fontsize{14}{17}\bfseries}}
    \renewcommand\subsubsection{\@startsection{subsubsection}{3}{\z@}%
                                        {-3.25ex\@plus -1ex \@minus -.2ex}%
                                         {1.5ex \@plus .2ex}%
                                         {\normalfont\normalsize\fontfamily{phv}\fontsize{14}{17}\selectfont}}
    \makeatother
	
	\usepackage{amsmath}
	\usepackage{graphicx}
	\usepackage{enumerate}
	\usepackage{amssymb}
	\usepackage{natbib} 
	\usepackage[hidelinks,colorlinks=true,linkcolor=blue,citecolor=blue,urlcolor=blue]{hyperref}
	\usepackage{xurl} 
	
        \graphicspath{{figs/}}
        \newcommand\extrafootertext[1]{%
        \bgroup
        \renewcommand\thefootnote{\fnsymbol{footnote}}%
        \renewcommand\thempfootnote{\fnsymbol{mpfootnote}}%
        \footnotetext[0]{#1}%
        \egroup
        }
	
\begin{document}

		\def\spacingset#1{\renewcommand{\baselinestretch}%
			{#1}\small\normalsize} \spacingset{1}
		\title{Decomposing LIBOR in Transition: Evidence from the Futures Markets}
		\date{}
		\if0\blind
		{
			\author{Jacob Bjerre Skov$^\star$ and David Skovmand$^\star$ \\
			$^\star$Department of Mathematics,\\ University of Copenhagen,\\ 2100 Copenhagen, Denmark.}
			\maketitle
		} \fi
		
		\if1\blind
		{
			\author{Author information is purposely removed for double-blind review}
			\maketitle
		} \fi
		\bigskip
	\spacingset{1.5} 
	\begin{abstract}
Applying historical data from the USD LIBOR transition period, we estimate a joint model for SOFR, Federal Funds, and Eurodollar futures rates as well as spot USD LIBOR and term repo rates. The framework endogenously models basis spreads between each of the benchmark rates and allows for the decomposition of spreads. Modelling the LIBOR-OIS spread as credit and funding-liquidity roll-over risk, we find that the spike in the LIBOR-OIS spread during the onset of COVID-19 was mainly due to credit risk, while on average credit and funding-liquidity risk contribute equally to the spread. 

	\end{abstract}
			
	\noindent%

	\noindent%

	\spacingset{1.5} 

\section{Introduction} \label{s:intro}
Fixed income markets are currently undergoing a major transition from the well established IBOR rates to overnight transaction-based rates termed Risk-Free rates (RFR) as the primary interest rate benchmark. In the US the Secured Overnight Financing Rate (SOFR) is scheduled to replace USD LIBOR by the middle of 2023. In particular, when facing a transition from LIBOR to SOFR, understanding and modelling what sets LIBOR apart from from SOFR from an empirical point of view is of great importance to market participants. Furthermore, the end of LIBOR has already prompted a wide variety of potential replacements,\footnote{For example the ICE Bank Yield Index (IBYI), Bloomberg Short-Term Bank Yield Index (BSBY), AMERIBOR, and AXI (see \cite{berndt2020across})} as SOFR, being a secured overnight rate, does not measure the actual cost of unsecured borrowing at term. In fact, \cite{klingler2021life} show that term rates based on SOFR can be detached from banks marginal funding and highlight the possible problematic implications of this for products such as credit lines during market stress. The introduction of several alternative rates and the ensuing discussions (see for example \cite{Risk1}), reflect the demand for a rate that captures the actual cost of funding as well as the need to understand and quantify the drivers of the funding costs.

Vast amounts of research in the field of term structure modelling have been devoted to the decomposition of LIBOR. Early studies such as \cite{collin2001term}, \cite{feldhutter2008decomposing}, and \cite{liu2006market} apply affine arbitrage-free multifactor models to study the impact of liquidity and credit risk on spread between LIBOR, interest rate swap rates, and treasury yields. The large increase in the spread between LIBOR and a maturity-matched overnight indexed swap (OIS) referencing the Effective Federal Funds Rate (EFFR) during the Great Financial Crisis further brought attention to the drivers of the LIBOR-OIS spread. \cite{michaud2008drives} argue that the lack of relationship between default risk and money market risk premia as well as the impact of central bank liquidity facilities on interbank rates indicate that liquidity is a key component to the size of the spread. \cite{dubecq2016credit} use a quadratic term structure model decompose the EURIBOR-OIS into credit and liquidity risks and evaluate effects of unconventional monetary policy in the euro zone during the crisis. \cite{filipovic2013term} study the decomposition of the LIBOR-OIS and EURIBOR-OIS spread during the Financial Crisis using credit default swap (CDS) data on LIBOR panel banks to identify a default and non-default component. Similarly, \cite{gallitschke2017interbank} study the decomposition of the LIBOR-OIS into credit and liquidity components using an equilibrium style modelling approach. \cite{backwell2019term} use similar data to decompose the EURIBOR-OIS spread into roll-over risk components. These studies all base their estimation of the LIBOR-OIS spread around swap data. 

In this paper we take a different approach and decompose the LIBOR-OIS spread by leveraging the fact that since the introduction of the SOFR futures contract in May of 2018 futures contracts referencing EFFR, SOFR, and LIBOR have traded simultaneously at the Chicago Mercantile Exchange (CME). This allows us to describe the dynamics of the benchmark rates using a joint model for Federal Funds, SOFR, and Eurodollar futures. While the majority of the existing research has focused on the decomposition of the LIBOR-OIS spread during the Financial Crisis, our paper, as a result of the fairly recent introduction of SOFR futures, studies the behaviour of the spread in the current transitory interest rate environment and in particular around the onset of the COVID-19 pandemic.

When modelling futures rates, we have to account for the futures convexity adjustment, see e.g. \cite{skov2021dynamic} and \cite{mercurio2018simple}. This is a model-dependent task, we therefore consider an affine framework for the the joint benchmark dynamics. Modelling term rates, we follow the approach of \cite{alfeus2020consistent} and \cite{backwell2019term} and model the LIBOR-OIS spread as a result of roll-over risk. By roll-over risk, we refer to the risk that an entity at a future point in time is not able to roll-over its loan at the prevailing reference rate, due to changes in credit or funding-liquidity risk. An increase in the roll-over risk components of our model thus leads to a steepening in the term structure of money market spreads consistent with the observations in \cite{gorton2021flight} showing that there was a "flight from maturity" during the Great Financial Crisis. 

We apply the Kalman filter quasi-maximum likelihood method to estimate the model to the historical record of spot and futures rates with up to one year of maturity. The model fit shows that the model is able to capture the cross-sectional and time variation in the term structure of futures rates for all three benchmark rates. As an additional validity check we show that the model is able to provide close fits to out of sample swap rates on OIS contracts for EFFR and SOFR as well as LIBOR swap contracts. Examining the latent state variables filtered by the model estimation indicates that the FOMC announcement on October 11th 2019 in response to the SOFR surge during September 2019 was effective in bringing down both the repo specific liquidity premium as well as the funding-liquidity component.

Finally, including data on term repos allow us to decompose the three- and six-month LIBOR-OIS spread into a credit and funding-liquidity component. Our framework suggests that the credit and funding components on average each contribute equally to the spread during our sample period. We also find that the large spike in the LIBOR-OIS spread during the outbreak of the COVID-19 crisis was mainly driven by an increase in credit risk. 

The paper is structured as follows. In section \ref{model_theory} we present the joint affine setup for SOFR, EFFR, LIBOR, and term repos as well as the main formulas required to compute spot and futures rates. Section \ref{Data_estimation} details the data and quasi-maximum likelihood method used to estimate the model. In section \ref{results} we discuss the empirical results.

\section{Constructing a Joint Interest Rate Setup for SOFR, EFFR, Term Repo, and LIBOR}\label{model_theory}
\subsection{Modelling SOFR and EFFR}
We consider a single risk-free short-rate, $r(t)$, defined on the filtered probability space $(\Omega,\mathcal{G},\{\mathcal{G}\}_{t\ge 0},Q)$. The risk-neutral measure, $Q$, is defined by the risk-free continuous savings account numeraire with value process $B(t)=B(0)e^{\int_0^t r(u) du}$.
The associated risk-free zero coupon bond price process is given by
\begin{equation}
    p(t,T)=B(t) \mathbb{E}^Q \left[B(T)^{-1} |\mathcal{F}(t)\right].
\end{equation}
The spread between the risk-free rate and SOFR is represented by a non-negative process $\psi(t)$. The process captures the systemic specific risk premia in the repo market such as gap risk, induced by varying liquidity of treasury securities, default risk in treasuries, haircuts etc. (see \cite{lou2020sofr} for an depth discussion of the magnitude of these particular risks and \cite{hu2021tri} for an analysis of pricing in the tri-party repo market). The SOFR specific short rate process is therefore given by
\begin{equation}
    r^s(t)=r(t)+\psi(t).\label{sofrshortrate}
\end{equation}
For identification reasons we posit dynamics for the SOFR related short rate process $r^s(t)$ directly using a two-factor Gaussian process
\begin{align}
&dr^s(t) = \kappa^r(\theta^s(t) - r^s(t)) dt + \sigma^r d W^r(t),\\
&d\theta^s(t) = \kappa^\theta(\theta^\theta - \theta^s(t)) dt + \sigma^\theta \left(\rho d W^r(t)+\sqrt{1-\rho^2}d W^\theta(t)\right). 
\end{align}
\cite{gellert2021short} and \cite{andersen2020spike} present modelling frameworks consistent with the spikes observed in overnight SOFR, however, as shown in \cite{skov2021dynamic} a simple Gaussian setup without jumps is sufficient to obtain a decent fit when modelling futures contracts based on compounded averages of the overnight benchmark. 

The spread between EFFR and the true risk-free rate is represented by an average overnight credit spread reflecting that the EFFR is an unsecured overnight rate. We denote the non-negative overnight credit spread, $\Lambda(t)$, and the affected short rate
\begin{equation}
    r^{FF}(t)=r(t)+\Lambda(t).
\end{equation}
$\Lambda(t)$ thus represents the time $t$ overnight credit spread of an average institution able to borrow at the EFFR. 

Consider the spread between EFFR and SOFR defined as 
\begin{equation}
\zeta(t):=r^{FF}(t)-r^s(t)= \Lambda(t)-\psi(t).
\end{equation} 
While both the overnight credit component and systemic repo component are non-negative, the spread between the two can turn both positive and negative depending on the size of each component. Furthermore, despite SOFR and EFFR represent a secure and unsecured rate, respectively, it is clear from the data and as we show in this paper that EFFR is not consistently greater than SOFR. The EFFR represents the unsecured rate in the funding market between top tier financial institutions that post reserves at the Fed, and thus approximately reflect the average cost of unsecured overnight funding of US LIBOR panel banks.\footnote{We note that the participants in the Federal Funds market are not limited to the US LIBOR panel banks, but consist of a larger set of institutions with accounts at the Federal Reserve banks. This includes US commercial banks, US branches of foreign banks, savings and loan organizations, and government-sponsored enterprises. The EFFR therefore only approximately reflects the cost of overnight unsecured funding of the panel banks.} On the other hand SOFR is derived from repo transactions and thus contains many idiosyncratic aspects specific to the repo market, exemplified most prominently in the SOFR Surge of September 2019. See \cite{lou2020sofr} for an in depth description of the repo market underlying SOFR. As a consequence we model the spread between EFFR and SOFR using a Gaussian process
\begin{equation}
d\zeta(t) = \kappa^\zeta(\theta^\zeta - \zeta(t)) dt + \sigma^\zeta d W^\zeta(t).    
\end{equation}
Modelling the spread above as opposed to the individual three components $r(t)$, $\Lambda(t)$, and $\psi(t)$ avoids the need of identifying the pure risk free rate $r(t)$, since a proper proxy for this rate is not available in the context of our model nor identified by our data. We stress that while we are unable to identify the individual components, our approach of directly modelling the SOFR specific short rate, $r^s(t)$, and the EFFR-SOFR spread process, $\zeta(t)$, still allow us to ascribe the correct underlying components to the equivalent secured and unsecured term rates as we describe in section \ref{section:termlibor} and \ref{section:TermRepo}.

\subsection{Modelling the LIBOR rate}
In modelling LIBOR we use a multi-curve model construction. The literature on multi-curve models is vast, and we refer to \cite{grbac2015interest} for an overview of the literature. In particular, we follow a structure similar to \cite{alfeus2020consistent} and ascribe the spread between overnight and term unsecured borrowing to roll-over risk. This approach is similar in spirit to the so-called renewal approach in \cite{collin2001term}, which was also applied in a different context in \cite{filipovic2013term}. Thus, we consider a time $t$ representative member of the LIBOR panel able to finance itself at both the overnight unsecured rate, EFFR, as well as on a fixed term basis with LIBOR.
Roll-over risk then represents the risk of a time $t$ representative member not being able to roll-over its overnight unsecured financing at a time $u$ with $u>t$. 
We denote the time $u$ roll-over risk infinitesimal spread of a time $t$ representative LIBOR panel member by $\gamma_t(u)$, and thus the total unsecured funding rate of this representative entity at time $u\geq t$
\begin{equation} \label{fund_r_1}
    r_t^f(u)=r^{FF}(u)+\gamma_t(u).
\end{equation}
Since such an entity is assumed to be able to finance itself using both unsecured overnight and term borrowing at time $t$ we require that $\gamma_t(t)=0$. However, for any $u>t$ the entity fixed at time $t$ may no longer be representative of the LIBOR panel. As such there is a risk of the the event $\gamma_t(u)>0$ occurring. I.e a roll-over risk event where the entity is unable to refinance its debt at the current market benchmark rate. The representative entity we consider in our model is therefore not to be understood as fixed, but potentially changing for every time $t$ that is considered.

The roll-over risk spread, $\gamma_t(u)$, can be decomposed into two separate components, a credit-downgrade component and a funding-liquidity component. The credit-downgrade component denoted $\lambda_t(u)$ reflects the credit deterioration of the time $t$ representative entity compared to the time $u$ updated reference panel. The funding-liquidity component captures the risk that the entity is not able to roll-over its debt at the reference rate without it being due to a decrease in credit quality. Such freezes in lending liquidity or the fear thereof are best associated with the Financial Crisis or perhaps more recently at the outbreak of the COVID crisis.
Thus, we model the total roll-over risk spread expressed as the sum of both credit-downgrade and funding-liquidity risk
\begin{equation}
    \gamma_t(u)=\lambda_t(u)+\phi_t(u).
\end{equation}
The rate which applies to funding an unsecured loan over the period $[t,T]$ using the unsecured overnight rate therefore contains both the overnight average credit spread as well as the roll-over risk specific components. Thus, for any time $u$ with $t\le u \le T$ the funding rate, $r_t^f(u)$, consists of 
\begin{equation} \label{fund_r}
    r_t^f(u)=r(u)+\Lambda(u)+\lambda_t(u)+\phi_t(u).
\end{equation}
Since the roll-over risk specific components are interpreted as future shocks in either credit quality or funding-liquidity and initiated at zero, we model these as pure jump processes. The credit component dynamics for $u\geq t$ are assumed to given as
\begin{equation}
d\lambda_t(u) = -\beta^\lambda \lambda_t(u) du +  d J^\lambda_t(u), \quad     \lambda_t(t)=0,
\end{equation}
and similarly the funding-liquidity component for $u\geq t$ is assumed to be given as
\begin{equation}
d\phi_t(u) = -\beta^\phi \phi_t(u) du +  d J^\phi_t(u), \quad     \phi_t(t)=0.
\end{equation}
In both cases the jump sizes are assumed exponentially distributed with a fixed mean of $2\%$. Furthermore, we will assume zero recovery at default throughout the paper. The recovery rate and mean jump size are interchangeable with the intensity level and thus these quantities are fixed in order to be able to identify the intensity processes.\footnote{Related studies such as \cite{filipovic2013term} and \cite{backwell2019term} follow a similar approach in order to identify the intensity process.}
The jumps in credit-downgrade component,  $J^\lambda_t(u)$, is assumed to have a stochastic intensity process modeled using a two-factor square-root process
\begin{align}
&d\xi(t) = \kappa^\xi(\eta(t) - \xi(t)) dt + \sigma^\xi \sqrt{\xi(t)} d W^\xi(t), \\
&d\eta(t) = \kappa^\eta(\theta^\eta - \eta(t)) dt + \sigma^\eta \sqrt{\eta(t)} d W^\eta(t).    
\end{align}
For the jumps, $J^\phi_t(u)$, in the funding-liquidity component, we apply a single square-root process
\begin{align}
d\nu(t) = \kappa^\nu(\theta^\nu - \nu(t)) dt + \sigma^\nu \sqrt{\nu(t)} d W^\nu(t).
\end{align}
As an identifying restriction (see the following section \ref{section:TermRepo}), we assume that 
$r(u)$, $\phi_t(u)$ and in turn $\nu(u)$ are independent of the underlying systemic repo risk process $\psi(u)$. The model is now fully specified under the risk neutral measure $Q$ and we can formulate the state vector process for $u\geq t$
$X_t(u):=(r^s(u),\theta^s(u),\zeta(u),\lambda_t(u),\phi_t(u),\xi(u),\eta(u),\nu(u))'$. As demonstrated in Appendix \ref{appendix:pricing} this defines an affine process in the sense of \cite{duffie2000transform} for each $t$, with dynamics 
\begin{equation}
    dX_t(u)=K^Q \left(\theta^Q-X_t(u) \right)du+\Sigma D(X_t(u),t) dW^Q(u)+dJ_t(u)
\end{equation}
the elements of which are explicitly defined in Appendix \ref{appendix:pricing}.
When estimating the model to the historical record of data, we also require the dynamics under the physical measure $P$.
The physical and risk-neutral measures are related through the market price of risk, $\mu(t)$, such that
\begin{equation}
    dW^Q(t)=dW^P(t)+\mu(t)dt.
\end{equation}
Since the components driving the roll-over risk are initiated at zero on each observation date we do not model their time series properties, and therefore only specify a market price of risk for the remaining state variables.
Estimating the drift specific parameters under $P$ is severely challenged by the fairly short sample of data on contracts referencing SOFR. We therefore consider a simple completely affine market price of risk structure given by
\begin{equation} \label{mpr}
  \mu(t)  =\left(\mu^r,\mu^\theta ,\mu^\zeta,\mu^\xi\sqrt{\xi(t)},\mu^\eta\sqrt{\eta(t)},\mu^\nu\sqrt{\nu(t)}\right)'.
\end{equation}
\subsection{Term Rates with Roll-Over Risk}
In this section we outline how to compute fair secured (repo) and unsecured (LIBOR) term rates. The fair term rates are determined by equating the present value of a term loan at the secured or unsecured term rate to the present value of a strategy in which the loan is rolled-over at the equivalent secured or unsecured overnight benchmark rate on an overnight, and in our model abstraction, continuous basis.

\subsubsection{Term LIBOR}\label{section:termlibor}
In order to compute the fair term LIBOR, we start by considering the present value of continuously rolling over a loan from time $t$ to time $T$ at the unsecured funding rate in \eqref{fund_r}
\begin{align}\label{U}
    U(t,T)&=
    B(t)E^Q\left[\frac{1}{B(T)}e^{\int_t^T r_t^f(u)du}
    1_{(\tau_t>T)} |\mathcal{G}_t\right] \nonumber\\
    &=B(t)E^Q\left[e^{\int_t^T \phi_t(u)+\lambda_t(u)+\Lambda(u) du} 1_{(\tau_t>T)} |\mathcal{G}_t\right] \nonumber\\
    &=B(t) E^Q\left[e^{\int_t^T \phi_t(u) du}|\mathcal{F}_t \right].
\end{align}
The last equality follows from the results on the intensity based credit risk approach in Appendix \ref{appendix:intensity}.
If instead the entity is able to borrow unsecured over the period $[t,T]$ at a rate $L(t,T)$, then the present value of the repayment is given by
\begin{align}
     B(t) E^Q\left[\frac{1}{B(T)}1_{(\tau_t>T)}\left(1+(T-t)L(t,T)\right)|\mathcal{G}_t \right].\label{termloan}
\end{align}
Next, we define the value of the defaultable zero coupon
\begin{align}
    Q(t,T)&=
    B(t) E^Q\left[\frac{1}{B(T)}1_{(\tau_t>T)}|\mathcal{G}_t \right] \nonumber\\
    &=B(t) E^Q\left[e^{-\int_t^T r(u)+\lambda_t(u)+\Lambda(u) du}|\mathcal{F}_t \right]
    \nonumber\\
    &=B(t) E^Q\left[e^{-\int_t^T r^s(u)+\zeta(u)+\lambda_t(u) du}|\mathcal{F}_t \right].
\end{align}
Where again we have used the intensity based credit risk approach as well as the relation $r(u)+\Lambda(u)=r^s(u)+\zeta(u)$ to obtain processes identified in the model setup.
The present value of the unsecured term loan in \eqref{termloan} can then be expressed as
\begin{align}
    \left(1+(T-t)L(t,T)\right)Q(t,T).\label{PV}
\end{align}
Since the value of the repayment from the roll-over risky account and the term loan must reflect the same present value, we equate \eqref{U} and \eqref{PV} to get the fair spot LIBOR rate
\begin{align} \label{libor_term}
    L(t,T)&=\frac{1}{T-t}\left(\frac{U(t,T)}{Q(t,T)} -1\right) \nonumber
    \\
    &=\frac{1}{T-t}\left(\frac{E^Q\left[e^{\int_t^T \phi_t(u) du}|\mathcal{F}_t \right]}{E^Q\left[e^{-\int_t^T r^s(u)+\zeta(u)+\lambda_t(u) du}|\mathcal{F}_t \right]} -1\right).
\end{align}
The term LIBOR is consistent with our interpretation of the EFFR as the rate at which a LIBOR panel bank is able to fund itself unsecured on a running basis. Indeed, as the term aspect vanishes the rate matches the rate of the overnight unsecured benchmark. Specifically, assuming differentiability with respect to $T$ and applying the definition $\lambda_t(t)=\phi_t(t)=0$
\begin{align}
     \underset{T\to t}\lim\ L(t,T)&=
     \frac{\partial}{\partial T}
     \left.\left(
     \frac{E^Q\left[e^{\int_t^T \phi_t(u) du}|\mathcal{F}_t \right]}{E^Q\left[e^{-\int_t^T r^s(u)+\zeta(u)+\lambda_t(u) du}|\mathcal{F}_t \right]}-1 
     \right)\right\rvert_{T = t} \nonumber\\
     &= \frac{\partial}{\partial T}
     \left.
     E^Q\left[e^{\int_t^T \phi_t(u) du}|\mathcal{F}_t \right]
     \right\rvert_{T = t}
     -
     \frac{\partial}{\partial T}
     \left.
     E^Q\left[e^{-\int_t^T r^s(u)+\zeta(u)+\lambda_t(u) du}|\mathcal{F}_t \right] 
     \right\rvert_{T = t}
     \nonumber\\
     &=r^s(t)+\zeta(t)+\lambda_t(t)+\phi_t(t)
    \nonumber\\
     &=r^{FF}(t).
\end{align}
The spot LIBOR is calculated using the affine model specifcation as
\begin{align}\label{libor_eq}
    L(t,T)=\frac{1}{T-t}\left(e^{A^{L}(T-t)+B^{L}(T-t)'X_t(t)}-1\right)
\end{align}
where $A^{L}(T-t)$ and $B^{L}(T-t)$ solve the spot LIBOR specific equations specified in Appendix \ref{appendix:pricing}.

\subsubsection{Term Repo Rate}
\label{section:TermRepo}
A repo loan, is a loan that is fully mitigated by credit risk since a secure underlying asset, i.e. a treasury instrument is posted as collateral with the counterparty. To determine the fair term Repo rate, $R^{repo}(t,T)$, we can thus follow a similar approach as before. A reasonable assumption is that the same entity that is a time-$t$ representative of the LIBOR panel also has access to a sufficiently large pool of treasury instruments and therefore is able to fund itself using repo transactions on a running, and in our model abstraction, continuous basis. This means it can fund itself at a continuous rate that disregards credit spreads altogether due to posting of the underlying asset with the counterparty. The funding rate of the entity on a continuous repo loan is therefore
\begin{equation} \label{fund_repo}
    r_t^{repo}(u)=r(u)+\psi(u)+\phi_t(u).
\end{equation}
Recall that $\psi(u)$ captures the systemic specific risk premia in the repo market, also present in the SOFR rate defined in Equation \eqref{sofrshortrate}. Similarly, $\phi_t(u)$ represents the funding-liquidity spread also present in the LIBOR rate. We first consider the present value of a repo strategy where one unit of currency is rolled-over at the continuous repo rate $r_t^{repo(u)}$ from $t$ to $T$. The time $t$ value is then given by
\begin{align} \label{secure_roll}
    B(t)E^Q\left[\frac{1}{B(T)}e^{\int_t^T r_t^{repo}(u)du}|\mathcal{F}_t \right]
    =E^Q\left[e^{\int_t^T\psi(u)+\phi_t(u)du}|\mathcal{F}_t \right].
\end{align}
Alternatively, the entity can borrow over the entire period using a term repo. The time $t$ value of the time $T$ repayment of the repo loan is
\begin{align}
    &B(t)E^Q\left[\frac{1}{B(T)}\left(1+(T-t)R^{repo}(t,T)\right)|\mathcal{F}_t \right] \nonumber\\
    &=B(t)E^Q\left[\frac{1}{B(T)}|\mathcal{F}_t \right]\left(1+(T-t)R^{repo}(t,T)\right).\label{pv2_repo}
\end{align}
Again, requiring equal present values of each approach by setting \eqref{secure_roll} equal to \eqref{pv2_repo} implies
\begin{equation}\label{repo_term}
    R^{repo}(t,T)=\frac{1}{T-t}\left( \frac{E^Q\left[e^{\int_t^T\psi(u)+\phi_t(u)du}|\mathcal{F}_t \right]}{E^Q\left[e^{-\int_t^T r(u) du}|\mathcal{F}_t \right]}-1\right).
\end{equation}
Analogously to the LIBOR case, one can show that our definition of the term repo rate is consistent with our interpretation of SOFR as the secured overnight funding rate in the sense that 
\begin{equation}
     \underset{T\to t}\lim\ R^{repo}(t,T)=r^{s}(t).
\end{equation}
Since the individual $r(u)$ and $\psi(u)$ processes are not identified in our model, we approximate the true term repo rate in \eqref{repo_term} to obtain an expression that can be calculated using processes identified by the model. The approximation assumes independence between $\psi(u)$ and $\phi_t(u)$ as well as $r(u)$ and $\psi(u)$. Furthermore, it ignores a convexity adjustment for $\psi(t)$ by applying the approximation $E^Q\left[\frac{1}{
    e^{\int_t^T\psi(u)du}}|\mathcal{F}_t \right]
    \approx
        \frac{1}{
    E^Q\left[e^{\int_t^T\psi(u)du}|\mathcal{F}_t \right]}$.
The resulting term repo approximation can then be calculated as
\begin{align}\label{repo_term_approx}
    R^{repo}(t,T)\approx
    \frac{1}{T-t}\left( \frac{E^Q\left[e^{\int_t^T\phi_t(u)du}|\mathcal{F}_t \right]}
    {E^Q\left[e^{-\int_t^T r^s(u) du}|\mathcal{F}_t \right]}-1\right).
\end{align}
We detail the derivation of the approximation in Appendix \ref{appendix:repo} and study its validity by creating an upper and lower bound for the true term repo rate. The bounds show that, given our model estimates, the resulting approximation error is less than $0.02$ and $0.08$ basis points for the three- and six-month repo contracts considered in the estimation. Again, the approximate term repo rate is calculated using the affine model specification
\begin{align}
    R^{repo}(t,T)&\approx\frac{1}{T-t}\left(E^Q\left[e^{\int_t^T r^s(u)+\phi_t(u)du}|\mathcal{F}_t \right]-1\right)\nonumber\\
    &=\frac{1}{T-t}\left(e^{A^{repo}(T-t)+B^{repo}(T-t)'X_t(t)}-1\right).
\end{align}
Where $A^{repo}(T-t)$ and $B^{repo}(T-t)$ solve the term repo specific equations (See Appendix \ref{appendix:pricing}).

\subsection{Futures Contracts}\label{futures_pricing}
Futures contracts traded at the CME are quoted as $100(1-R(S,T))$ where $R(S,T)$ denotes the contract specific futures rate which comes in the four different variants as described below. Following standard results (see e.g. \cite{hunt2004financial}) the value of a futures contract with a random payoff is given by the risk-neutral expectation of the non-discounted payoff. Applying this to interest rate futures the time $t$ futures rate is given by
\begin{equation}\label{futures_eq}
    f(t;S,T)=\mathbb{E}^Q\left[R(S,T)|\mathcal{F}_t\right].
\end{equation}
\subsubsection{Eurodollar Futures}\label{edfut}
Eurodollar futures contracts reference a future three-month LIBOR fixing. The Eurodollar futures rate is therefore given by
\begin{equation}
    f^{ED}(t;S,T)=\mathbb{E}^Q\left[L(S,T)|\mathcal{F}_t\right].
\end{equation}
Inserting the spot LIBOR expression in \eqref{libor_eq} we have
\begin{align}
    f^{ED}(t;S,T)
    &=\mathbb{E}^Q\left[
    \frac{1}{T-S}\left(E^Q\left[e^{\int_S^T r^s(u)+\zeta(u)+\lambda_S(u)+\phi_S(u) du}|\mathcal{F}_S \right]-1\right)|\mathcal{F}_t\right] \nonumber\\
    &=\frac{1}{T-S}\mathbb{E}^Q\left[
    \left(e^{A^L(T-S)+B^L(T-S)'X_S(S)}-1\right)|\mathcal{F}_t\right] \nonumber\\
    &=\frac{1}{T-S}\mathbb{E}^Q\left[
    \left(e^{A^L(T-S)+\hat{B}^L(T-S)'X_t(S)}-1\right)|\mathcal{F}_t\right].
\end{align}
In the last equation we recall that the jump specific elements, $\lambda_S(S)$ and $\phi_S(S)$ in $X_S(S)$ are zero by definition. Thus, we let $\hat{B}^L(T-S)$ be identical to $B^L(T-S)$, but with zeros in the jump specific elements. The Eurodollar futures rates is then calculated as
\begin{align}\label{eurodollar}
     f^{ED}(t;S,T)=\frac{1}{T-S}\left(e^{A^{ED}(S-t)+B^{ED}(S-t)'X_t(t)}-1\right).
\end{align}
Where $A^{ED}(S-t)$ and $B^{ED}(S-t)$ solve the Eurodollar futures specific Riccati equations (see Appendix \ref{appendix:pricing}) with initial conditions $A^{ED}(0)=A^L(T-S)$ and $B^{ED}(0)=\hat{B}^L(T-S)$ with $T-S=91/360$. 

\subsubsection{Three-Month SOFR Futures}\label{3msofrfut}
The three-month SOFR futures rate, $R^{s,3m}(S,T)$, is computed as the daily compounded SOFR fixings during the reference quarter
\begin{equation}\label{sofr3mfut}
    R^{s,3m}(S,T)=\frac{1}{T-S}\left(\prod_{i=1}^N \left(1 +d_i R_{d_i}^{s}(t_i) \right)-1 \right),
\end{equation}
with $R_{d_i}^{s}(t_i)$ for $i\in 1,...,N$ with $S\le t_1,...,t_N\le T$ denoting the realized SOFR fixings in the reference quarter and $d_{i}$ the day count fraction multiplied amount of days to which $R_{d_i}^{s}(t_i)$ applies.\footnote{E.g. for Fridays $d_{i}=3/360$ while $d_{i}=1/360$ on days with a normal business day the following day.} When computing the futures rate we consider the continuous approximation of the futures rate
\begin{equation}
R^{s,3m}(S,T)\approx \frac{1}{T-S}\left(e^{\int_S^T r^s(u) du}-1\right).
\end{equation}
The errors induced by this approximation are studied in detail in Appendix D in \cite{skov2021dynamic} and they are found to be of no economic significance. Given the approximation, the three-month SOFR futures rate can be computed as
\begin{align}
    f^{s,3m}(t;S,T)
    &=\mathbb{E}^Q\left[
    \frac{1}{T-S}\left(e^{\int_S^T r^s(u) du}-1\right)|\mathcal{F}_t\right] \nonumber\\
    &=\frac{1}{T-S}\mathbb{E}^Q\left[\mathbb{E}^Q\left[
    \left(e^{\int_S^T r^s(u) du}-1\right)|\mathcal{F}_S\right]|\mathcal{F}_t\right] \nonumber\\
    &=\frac{1}{T-S}\mathbb{E}^Q\left[
    \left(e^{A^{s,3m}(T-S)+B^{s,3m}(T-S)'X_S(S)}-1\right)|\mathcal{F}_t\right] \nonumber\\
    &=
    \frac{1}{T-S}\left(e^{A^{s,f}(S-t)+B^{s,f}(S-t)'X_t(t)}-1\right).
\end{align}
Where $A^{s,3m}(T-S)$, $B^{s,3m}(T-S)$, $A^{s,f}(S-t)$, and $B^{s,f}(S-t)$ all solve Riccati equations presented in Appendix \ref{appendix:pricing} with initial conditions $A^{s,3m}(0)=B^{s,3m}(0)=0$, $A^{s,f}(0)=A^{s,3m}(T-S)$, and $B^{s,f}(0)=B^{s,3m}(T-S)$ with $T-S=91/360$ reflecting the accrual days in the three-month SOFR futures contract.
It is important to emphasize that while the LIBOR fixing is a forward looking term rate and thus $\mathcal{F}_S$-measurable, the SOFR futures rate is based on a backward-looking compounded rate and therefore only $\mathcal{F}_T$-measurable. Thus, for $S<t<T$ we have to account for the part of the underlying that has already been accrued
\begin{align}
f^{s,3m}(t;S,T)=
\frac{1}{T-S}\left(\left(\prod_{i=1}^{N_0} \left[1+d_i R^{s}_{d_i}(t_i) \right]\right)e^{A^{s,3m}(T-t)+B^{s,3m}(T-t)'X_t(t)} - 1 \right)
\end{align}
where $R^s_{d_i}(t_i)$ for $i\in 1,...,N_0$ with $S\le t_1,...,t_{N_0}\le t$ are the $\mathcal{F}_t$-measurable realized SOFR fixings.

\subsubsection{One-Month SOFR Futures}
The one-month SOFR futures rate is based on the arithmetic average of the daily SOFR fixings during the contract month
\begin{align}
R^{s,1m}(S,T)=\frac{1}{T-S}\sum_{i=1}^N d_i R^{s}_{d_i}(t_i),
\end{align}
with $N$ the total number of days in the month and $R^{s,1m}_{d_i}(t_i)$ for $i\in 1,...,N$ with $S\le t_1,...,t_N\le T$ the published SOFR. For any date for which the rate is not published the last preceding rate is used. As in \cite{mercurio2018simple} we approximate the discrete average by an integral of the instantaneous short rate
\begin{equation}
R^{s,1m}(S,T)\approx \frac{1}{T-S}\int_S^T r^s(u) du.\label{approxarithmetiaverage}
\end{equation}
Again, the errors induced by this approximation are studied in detail in Appendix D in \cite{skov2021dynamic} and found to be of no economic significance. 
The one-month SOFR futures rate can therefore be computed as follows 
\begin{align} 
    f^{s,1m}(t;S,T)
    &=\mathbb{E}^Q\left[
    \frac{1}{T-S}\int_S^T r^s(u) du|\mathcal{F}_t\right].\label{sofr1m}
\end{align}
The expectation is calculated explicitly in Appendix \ref{appendix:pricing}.
For spot contracts when $S<t<T$ a part of the total futures rate has already fixed, which we account for by letting
\begin{align} \label{sofr1m2}
f^{s,1m}(t;S,T)
=\frac{1}{T-S} \sum_{i=1}^{N_0} d_i R^{s}_{d_i}(t_i)
+\mathbb{E}^Q\left[
    \frac{1}{T-S}\int_t^T r^s(u) du|\mathcal{F}_t\right]
\end{align}
where $R^{s}_{d_i}(t_i)$ for $i\in 1,...,N_0$ with $S\le t_1,...,t_{N_0}\le t$ are the realized SOFR fixings.

\subsubsection{Federal Funds Futures}
The federal funds futures contract has a one-month reference period and shares the same specifications as the one-month SOFR futures contract, however referencing the daily EFFR fixings. The approach to deriving the futures rate is therefore analogous to the previous section, and we get for $t\leq S<T$
\begin{align}  
    f^{FF}(t;S,T)
    &=\mathbb{E}^Q\left[
    \frac{1}{T-S}\int_S^T r^{FF}(u) du|\mathcal{F}_t\right].\label{ff1}
\end{align}
Again, we refer to Appendix \ref{appendix:pricing} for an explicit expression of the expectation.
\section{Data and Estimation}\label{Data_estimation}
Our dataset is collected through Refinitiv. It consists of daily observations on spot three- and six-month LIBOR fixings, the three- and six-month term repo for treasuries, Eurodollar futures, one- and three-month SOFR futures, and federal funds futures. Liquidity in the interest rate futures markets is generally concentrated in the contracts closest to expiry. Since the focus of this study is also the short end of the term structure, we include contracts with up to around one year of maturity.\footnote{Since futures contracts have fixed expiration dates the time to maturity of the included contracts will vary across different observation dates in our sample.} Volume in the futures market for SOFR contracts at CME has steadily increased since its beginning in May 2018. \cite{heitfield2019inferring} present data for the first year of trading in SOFR futures showing liquidity in the five nearest one-month contracts and nine nearest three-month contracts. Following this we include the five nearest one-month and five nearest three-month SOFR futures contracts in the data sample. The market for Eurodollar futures is massive and contracts are liquid out to a maturity of five years, however, in order to obtain an identical range to the SOFR contracts we include the four nearest Eurodollar contracts. Federal funds futures contracts are traded one and a half years out. We include the 12 nearest contracts in our estimation.\footnote{Daily volumes on all futures contracts can be found at \url{https://www.cmegroup.com/}}
Our data sample begins in June 2018 when SOFR futures started trading at the CME and runs until October 2021 resulting in a total of 840 daily observations. 

Based on the futures contracts included in the estimation, we note that our model estimation covers the term structure of the benchmark rates out to maturities of approximately one year. In this study, we thus solely focus on the short-term joint dynamics of the interest rate benchmarks. The short-term focus is unavoidable since our study relies on the coexistence of derivatives referencing actual independent fixings for both SOFR and LIBOR which are only available during the transition period. Furthermore, the cessation of US LIBOR was initially announced to happen by the end of 2021, this was then extended on November 30, 2020 to June 30, 2023 for all maturities except the one-week and two-month LIBOR. Therefore, the implied rate by any derivative referencing LIBOR traded before November 30, 2020 and maturing after 2021 as well as any derivative traded after November 30, 2020 and maturing after June 2023 will reflect the LIBOR fallback language rather than the actual cost of interbank funding (see section \ref{risk_prem} for further details on the Eurodollar futures LIBOR fallback methodology).

There are several reasons for looking at the futures market when modelling the short end of the term structure. First, the futures market is by far the most liquid short term market referencing the benchmark rates. Second, a historical record of contracts referencing SOFR is only available in the futures market. Finally, using futures contracts we do not have to take into account the discounting rate (see Equaion \ref{futures_eq}). This is especially relevant during our data sample since the price alignment interest (PAI) used by LCH and CME in e.g. swaps changed in October 2020 from EFFR to SOFR.

When estimating the model we transform the observed spot LIBOR rates into equivalent yields
\begin{equation}
    y^L(t,T)=\frac{1}{T-t}\log \left(1+(T-t)L(t,T) \right)
\end{equation}
As with the spot LIBOR rates we also transform the term repo
\begin{equation}
    y^{repo}(t,T)=\frac{1}{T-t}\log \left(1+(T-t)R^{repo}(t,T) \right).
\end{equation}
Following \cite{bikbov2005term} we also consider a similar transformation to obtain Eurodollar futures yields
\begin{equation}
    y^{ED}(t;S,T)=\frac{1}{T-S}\log \left(1+(T-S)f^{ED}(t;S,T) \right),
\end{equation}
as well as three-month SOFR futures yields
\begin{equation}
    y^{s,3m}(t;S,T)=\frac{1}{T-S}\log \left(1+(T-S)f^{s,3m}(t;S,T) \right).
\end{equation}
The transformations imply that the spot and futures rates are affine in the state variables and thus allow us to apply the standard Kalman filter when estimating the model (see Appendix \ref{appendix:kalman} for details on the Kalman filter). The federal funds and one-month SOFR futures rate approximations in Equation \eqref{ff1} and \eqref{sofr1m} are already affine in the state variables, therefore a similar transform is therefore not required for these contracts.

\section{Empirical Results}\label{results}
In this section we examine the consistency of the model outputs with observed data as well as estimates and latent variables. We then proceed to use our framework to decompose the spot three-and six-month LIBOR and compare risk premia in the futures markets.
\subsection{Model Fit}
\begin{figure}[hbt!]
\centering
\includegraphics[angle=0, scale=.8]{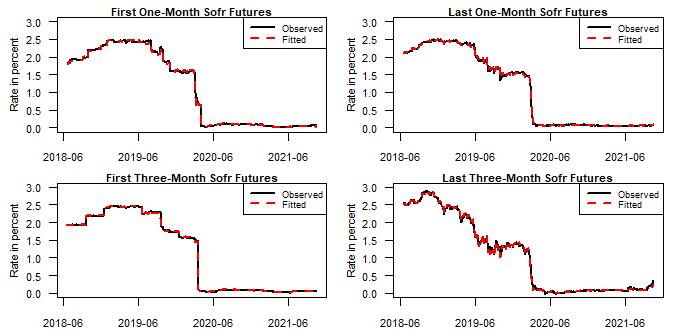}
\caption{In sample fit of the first and last one- and three-month SOFR futures contracts.}
\label{fig:sofr_fut}
\end{figure}

To validate that the model is able to capture the variation in futures and spot rates across our sample period, we compare the fitted values to the observed values. Figure \ref{fig:sofr_fut} plots the futures rates for a subset of the SOFR contracts in our data sample. The futures expiry dates are fixed dates, thus the time to maturity for each of the plotted contracts vary through the sample. E.g. the expiry of the nearest SOFR futures contract varies from one day to three months, while the second nearest SOFR futures contract has an expiry range between three and six months and so on. The resulting fit reflects the results presented in \cite{skov2021dynamic} showing that two Gaussian factors are enough to capture the majority of the variation in the SOFR futures market up to a one year maturity.

\begin{figure}[hbt!]
\centering
\includegraphics[angle=0, scale=.8]{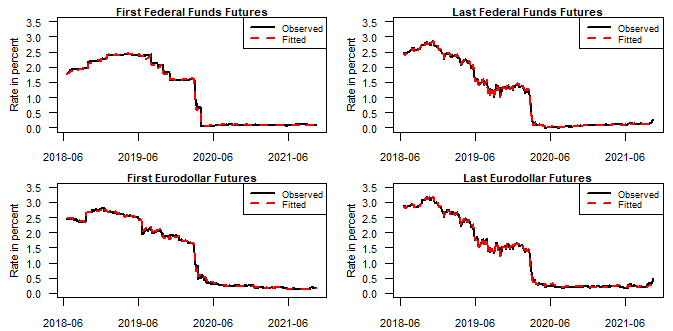}
\caption{In sample fit of the first and last federal funds and Eurodollar futures contracts}
\label{fig:fed_ed_fut}
\end{figure}

\begin{figure}[hbt!]
\centering
\includegraphics[angle=0, scale=.8]{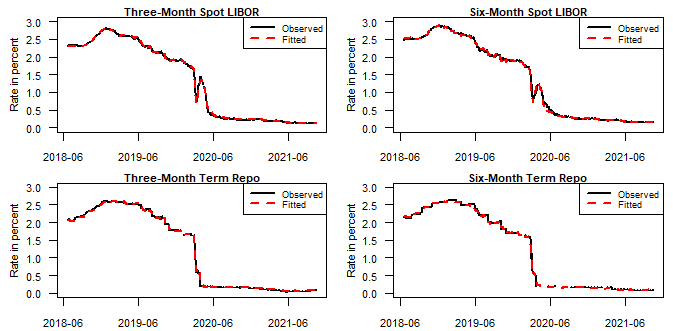}
\caption{In sample fit of three- and six-month spot LIBOR and repo rates.}
\label{fig:libor_repo}
\end{figure}

The Eurodollar and federal funds futures rates are plotted in figure \ref{fig:fed_ed_fut}. Focusing on the spot Eurodollar futures rate, we note the spike during the onset of the COVID-19 crisis reflecting the spike in spot LIBOR as seen in figure \ref{fig:libor_repo}. However, the increase is far less pronounced in the later contracts indicating that the Eurodollar futures market predicted the spike in LIBOR to be fairly short-lived.

In addition to futures contracts, we also include spot LIBOR and term repo in the estimation. The fit of these rates is plotted in Figure \ref{fig:libor_repo}. We see that the roll-over risk approach is able to capture the large spike in the LIBOR while also matching the term repo rate during the period in March 2020. We also note that term repos are fairly illiquid contracts, which is clearly seen from the observed six-month term repo data series showing multiple missing values, particularly during the market stress just after the onset of the COVID crisis. However, missing values are easily overcome during the filtering used in the estimation (see Appendix \ref{appendix:kalman}).

Table \ref{tab:RMSE} shows the fitted root mean squared errors (RMSE). When computing the RMSEs, we consider the actual futures and spot rates instead of the yield transformations used in the estimation. The RMSEs further demonstrate that the model provides a close fit across futures contracts on all three benchmarks.

\begin{table}[hbt!]
\centering
\begin{tabular}{cccccc}
     & SOFR Futures & EFFR Futures & ED Futures & Spot LIBOR & Term Repo \\ \hline
RMSE & 2.2          & 2.0          & 2.9        & 2.5      & 2.4      \\ \hline
\end{tabular}
\label{tab:RMSE}
\caption{RMSEs of the fitted rates for the full sample. All values are in basis points.}
\end{table}

\subsection{Estimates}
Table \ref{table:estimates} presents the estimated parameter values and standard deviations. First, we note that the process governing the spread between the overnight rates SOFR and EFFR has an estimated mean level, $\theta^\zeta$, equal to zero together with a low volatility, $\sigma^\zeta$, reflecting the high level of correlation between SOFR and the effective federal funds rate. Turning to the roll-over risk specific estimates, we see that the intensity process is rather volatile, however, with a quick mean reversion, while the stochastic mean process is more stable as also seen from the state variables in figure \ref{fig:states}. The large estimate of $\beta^\phi$ indicates that the funding-liquidity specific shocks are expected to quickly revert back to a normal state again. The parameters related to the market price of risk are estimated with significant standard errors, however, this is as expected given the fairly short sample size.
\begin{table}[hbt!]
\centering
\small
\noindent\makebox[\textwidth]{
\begin{tabular}{cccccccc}
Mean Reversion   & Estimate & Mean             & Estimate & Volatility       & Estimate & Risk Premium  & Estimate  \\ \hline
$\kappa^r$       & 1.2394   &                  &          & $\sigma^r$       & 0.0032   & $\mu^r$       & -1.3117   \\
                 & (0.0039) &                  &          &                  & (0.0000) &               & (1.4424)  \\
$\kappa^\theta$  & 0.0273   & $\theta^\theta$  & 0.0306   & $\sigma^\theta$  & 0.0071   & $\mu^\theta$  & 0.0003    \\
                 & (0.0017) &                  & (0.0012) &                  & (0.0001) &               & (0.5767)  \\
$\kappa^\zeta$   & 0.5945   & $\theta^\zeta$   & 0.0000   & $\sigma^\zeta$   & 0.0006   & $\mu^\zeta$   & -0.1095   \\
                 & (0.0808) &                  & (0.0000) &                  & (0.0001) &               & (1.0250) \\
$\kappa^\xi$     & 8.2375   &                  &          & $\sigma^\xi$     & 2.8610   & $\mu^\xi$     & 0.8202    \\
                 & (0.1714) &                  &          &                  & (0.1607) &               & (1.1110)  \\
$\kappa^\eta$    & 0.1299   & $\theta^\eta$    & 0.0163   & $\sigma^\eta$    & 0.7715   & $\mu^\eta$    & 0.1451    \\
                 & (0.0405) &                  & (0.3187) &                  & (0.0571) &               & (0.4502)  \\
$\kappa^\nu$    & 1.6624   & $\theta^\nu$    & 2.4408   & $\sigma^\nu$    & 3.1921   & $\mu^\nu$    & -0.2445    \\
                 & (0.0651) &                  & (0.3218) &                  & (0.2297) &               & (0.4180)  \\
$\beta^\lambda$   & 5.1952  &                  &          &                  &          &               &           \\
                 & (0.0742) &                  &          &                  &          &               &           \\
$\beta^\phi$   & 37.3898  &                  &          &                  &          &               &           \\
                 & (3.9829) &                  &          &                  &          &               &           \\ \hline
\end{tabular}
}
\caption{Parameter estimates with standard deviations in parentheses. The correlation parameter, $\rho$, is estimated at $0.0650 \  (0.0276)$. The estimated values of the filtering parameters and their standard deviations in basis points are $\sigma_{SOFR}=2.3094\ (0.0139)$,
$\sigma_{EFFR}=2.0621\ (0.0095)$, and 
$\sigma_{LIBOR}=2.8949\ (0.0164)$. The maximized log-likelihood value is $170,331$.}
\label{table:estimates}
\end{table}
\subsection{State Variables}
\begin{figure}[hbt!]
\centering
\includegraphics[angle=0, scale=.8]{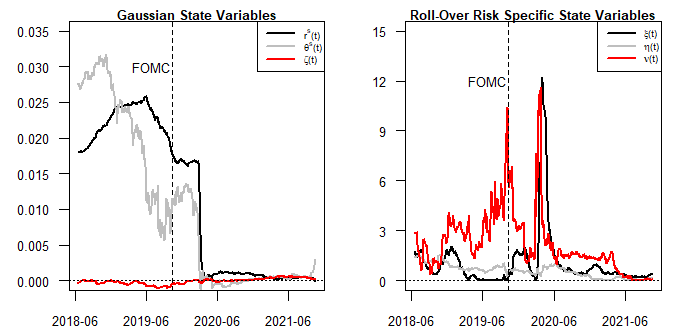}
\caption{The plot displays the filtered state variables impacting to the overnight benchmark rates during the sample period from June 2018 to October 2021. The vertical line marks the October 11th FOMC meeting.}
\label{fig:states}
\end{figure}

To investigate the drivers of the interest rate benchmarks, we plot the path of the filtered state variables during the data period in figure \ref{fig:states}. The period reflects a decrease in short term interest rate expectations as seen by the downward trend of the stochastic mean $\theta^s(t)$, which even turns slightly negative during the first half of 2020. The $\zeta(t)$ process related to the spread between the overnight benchmarks, SOFR and EFFR, is concentrated around zero reflecting both positive and negative spreads between the two rates. During the SOFR surge on September 17, SOFR increased to above $5$ percent.\footnote{See \cite{anbil2020happened} for details on what happened during the SOFR surge. \cite{copeland2021reserves} argue that the spikes in SOFR were a result of the reduced aggregate reserves following the balance-sheet normalization between 2017 and September 2019.} We note that the spread process also turns negative during the same period indicating an increase in the underlying systemic repo risk component. The decomposition thus shows that neither of the underlying spreads, $\psi(t)$ and $\Lambda(t)$, dominate the other one during the entire sample. This further emphasizes that even though SOFR is a secured rate unlike the EFFR, SOFR should not be thought of as being closer to a "pure" risk free rate than the EFFR. 
Turning to the roll-over risk specific variables represented by the jump intensity processes in the instantaneous funding rate. The process $\xi(t)$ relating to the credit downgrade component with its stochastic drift term $\eta(t)$ and $\nu(t)$ the intensity of jumps in the funding-liquidity spread are plotted in the right hand panel of figure \ref{fig:states}. In the period up to the SOFR surge of September 2019 the model identifies a general liquidity-spread increase through the rise in the intensity process $\nu(t)$, as well as an increase in SOFR relative to EFFR as the $\zeta(t)$ process moves into negative territory. All the while the credit-downgrade intensity $\xi(t)$ hovers around zero in the same period. 
On October 11th 2019 following the SOFR surge the fed announced that it would conduct operations in the term and overnight repo market.\footnote{See \url{https://www.federalreserve.gov/newsevents/pressreleases/monetary20191011a.htm}} After the announcement, we see an immediate increase in the SOFR-EFFR spread component suggesting a decrease in the systemic repo risk premium. Furthermore, the funding-liquidity component also decreases significantly. The decomposition therefore indicates that the actions of the fed did help to decrease risk premia both in the overnight and term rate market.\footnote{\cite{allen2009interbank} create a model for interbank liquidity and the intervention of central banks. Likewise, \cite{christensen2014central} find that the fed's Term Auction Facility (TAF) helped to reduce risk premia in interbank lending rates during the Financial Crisis.}

The Covid-19 related events of early March 2020 produce a massive spread between term LIBOR and the federal funds rate. Figure \ref{fig:states} shows that the increase in risk was, unlike the events following the SOFR surge, not just isolated to funding-liquidity risk. It affected the average overnight credit spread, as well as both the credit-downgrade and funding-liquidity risk. This can be seen in the clear spikes in both intensity processes $\nu(t)$, and $\xi(t)$ during the same period. Likewise, the low LIBOR-OIS spread in the second half of 2021 results in close to zero intensity processes.
\subsection{Comparing with Swap Contracts}\label{swap_comp}
As an additional validity check of the model, we perform an out of sample test of the model by comparing short-term observed swap rates with model implied rates. We detail the pricing of OIS and IRS contracts in Appendix \ref{appendix:pricing}. 

\begin{figure}[hbt!]
\centering
\includegraphics[angle=0, scale=.8]{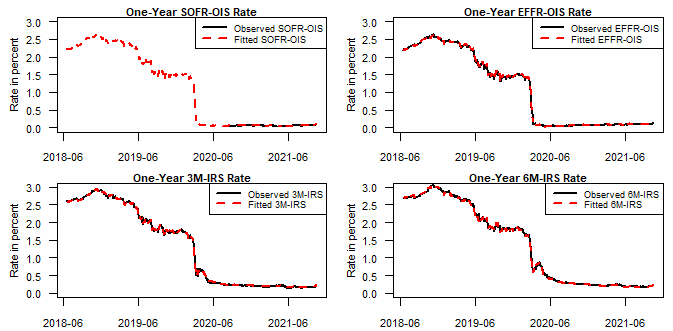}
\caption{Observed and model implied one-year swap rates. SOFR swap data is only available from July 2020.}
\label{fig:swap_plots}
\end{figure}

The model endogenously models the spread between different LIBOR tenors as a result of roll-over risk. Therefore, we also test the ability of the model to extrapolate the spot six-month rate to swaps of greater maturity referencing the six-month LIBOR fixing. To obtain data on the six-month LIBOR IRS we use that the basis swap rate, $BS^{3M,6M}(t,T)$, reflects the difference in fixed rates of two IRS referencing different LIBOR tenors. Swap rates referencing the six-month LIBOR fixing are then calculated as
\begin{equation}
    IRS^{6M}(t,T)=IRS^{3M}(t,T)+BS^{3M,6M}(t,T).
\end{equation}
Figure \ref{fig:swap_plots} and table \ref{tab:RMSE_swap} displays the fitted swap rates along with their RMSEs, and shows a tight fit reflecting a swap and futures market that is well in line with each other. Swap data could of course be added to the estimation data, likely resulting in a closer fit to the swap market. Furthermore, the swap market is traded at far greater maturities compared to futures thus allowing for modelling of the mid to long term maturities of the term structure as in \cite{filipovic2013term} and \cite{backwell2019term}. We restrict the scope of our estimation to futures contracts due to the limited liquidity of the SOFR swap market during our sample period. But also, as discussed in section \ref{Data_estimation}, because LIBOR derivatives expiring after June 2023 will reference SOFR plus the ISDA fixed fallback spread and thus not reflect actual expectations of interbank funding.

\begin{table}[hbt!]
\centering
\begin{tabular}{ccccc}
Maturity & SOFR-OIS & EFFR-OIS & 3M-IRS & 6M-IRS \\ \hline
1Y & 0.7          & 0.9          & 1.4          & 1.9     \\ \hline
\end{tabular}
\caption{RMSEs of the fitted swap rates for the full sample. All values are in basis points. SOFR swap data is only available from July 2020.}
\label{tab:RMSE_swap}
\end{table}

When decomposing the LIBOR-OIS spread into its roll-over risk components it is essential that the model correctly ascribes credit risk in the LIBOR panel to the roll-over risk credit component $\lambda_t(u)$. We investigate the decomposition by comparing the model-implied CDS-spread for an average LIBOR panel bank to the observed CDS spreads for the LIBOR panel banks. In particular, we construct the time series of CDS-spreads by trimming and averaging CDS-quotes for all banks in the current USD LIBOR panel using the LIBOR specific methodology.\footnote{The current USD LIBOR methodology and panel banks can be found at \url{https://www.theice.com/iba/libor}} We focus on the the shortest on-the-run CDS contract, which has a six month tenor, however, due to the biannual roll dates for these contracts the effective maturity varies between six months and a full year. Since we cannot identify the overnight average credit spread $\Lambda(t)$ using our dataset, we are only able to calculate the fair CDS-spread implied from the roll-over risk specific credit component. We detail the calculation of the fair swap spread in Appendix \ref{appendix:pricing}. Figure \ref{fig:cds} plots the model implied spread against the observed LIBOR panel CDS spread.
\begin{figure}[hbt!]
\centering
\includegraphics[angle=0, scale=.8]{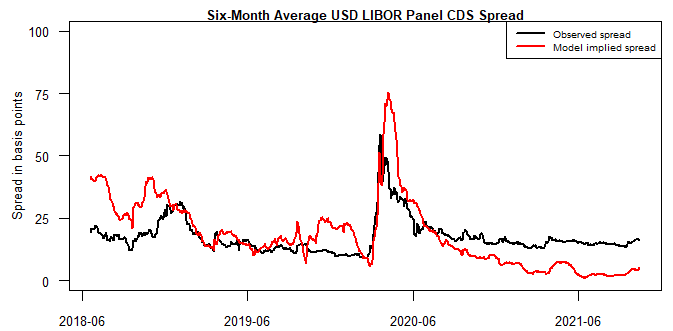}
\caption{Observed and model implied six month CDS spread based on the decomposed roll-over risk credit component.}
\label{fig:cds}
\end{figure}
In addition to excluding the overnight average credit component there are multiple reasons why we would not expect the model to produce an exact fit to the CDS spread. First, any idiosyncrasies in the short-term CDS market will not be reflected by the model fit without including these contracts in the estimation. Second, Since our estimation only includes data on the spot three- and six-month repo the decomposition of roll-over risk beyond the six-month tenor will be depend on an extrapolation. However, despite this the plot clearly still shows that the decomposed credit component captures variation in credit risk for the LIBOR panel banks. This is also apparent from the correlation of $75.2\%$ between changes in the monthly averages of the credit-downgrade intensity process, $\xi(t)$, and the observed average CDS spread. 

\subsection{Decomposing Roll-over Risk in Spot LIBOR} \label{decomposition}
In order to investigate the cost of unsecured interbank term funding present in LIBOR, we use our model to decompose the roll-over risk components in the spot three- and six-month LIBOR. As previously mentioned the LIBOR-OIS spread contains both credit and funding-liquidity risks, which cannot be separated using only data on EFFR and LIBOR. Previous studies such as \cite{filipovic2013term} and \cite{backwell2019term} use CDS data on LIBOR panel banks in order to identify the credit component and thus separate the roll-over risk specific components. These studies attempt to identify the entire term structure of the IRS-OIS spread. Due to the short term nature of this study, we focus solely on the decomposition of the three- and six-month spot LIBOR-OIS spread, however, CDS contracts are only traded at maturities of six months or longer with most of the liquidity concentrated in the five year maturity contract.\footnote{Due to the biannual roll dates for CDS contracts the actual contract maturities are usually greater than the quoted tenors. E.g. the actual expiry of the shortest CDS contract varies between six months and up to an entire year. See \cite{boyarchenko2019long} for details on the CDS market} To avoid this mismatch in maturities, we take a different approach and include data on the three- and six-month term repo rate. A similar method is used in \cite{smith2010term} who applies the three-month LIBOR-repo spread to identify the credit component in LIBOR. As we formalize in section \ref{section:TermRepo} the term repo rate is a secured rate and therefore unaffected by the overnight average credit spread as well as credit-downgrade risk. However, it is a term loan and thus still contains funding-liquidity risk. As an identification strategy our is constructed so that funding-liquidity risk impacts the LIBOR and term repo market equally through the same funding-liquidity risk spread $\phi_t(u)$. We note that in contrast to the previously mentioned studies in \cite{filipovic2013term} and \cite{backwell2019term} such a decomposition of the term repo is only applicable since we have included data on SOFR, allowing us to define the value of the overnight rolled-over secured loan in Equation \eqref{secure_roll} which accounts for the systemic repo risk premium captured by $\psi(t)$. Furthermore, this allows us to remain agnostic about the size of the average overnight credit risk level and in turn the actual underlying risk-free rate unlike previous studies where the overnight average credit spread has been assumed either as a fixed level of e.g. $5$ basis points (as in \cite{filipovic2013term} and \cite{alfeus2020consistent}) or extrapolated from credit default swap contracts (as in \cite{backwell2019term}).

\begin{figure}[hbt!]
\centering
\includegraphics[angle=0, scale=0.8]{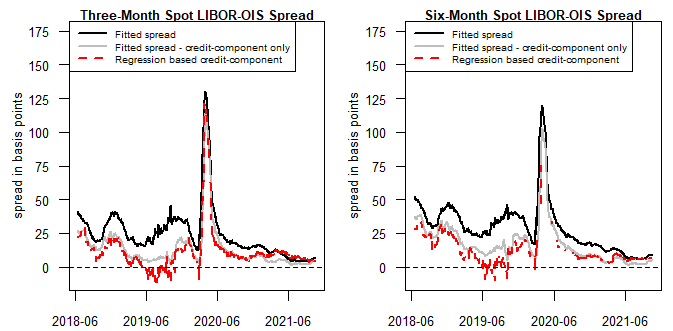}
\caption{Decomposed three- and six-month LIBOR-OIS spreads.}
\label{fig:roll_spread}
\end{figure}

Figure \ref{fig:roll_spread} plots the decomposed three- and six-month LIBOR-OIS spreads. First, we note that the large spike in the spread in March 2020 was mainly driven by a large increase in credit-downgrade risk. This observation also aligns with studies of the LIBOR-OIS spread during the great financial such as \cite{king2020credit} who find that at the peak of the crisis the spread was dominated by credit risk. During normal periods, however, we find that the funding component is equally relevant in explaining the spread. Considering the entire sample period the credit component explains $51.2\%$ of the three-month LIBOR-OIS spread and $53.9\%$ of the six-month spread. Comparing with similar studies, \cite{filipovic2013term} find that the impact of the non-default component is greater than the credit component for the spot three-month spread in both the USD and euro market. Similarly, \cite{schwarz2019mind} show that her liquidity measure dominates the impact of credit on the EURIBOR-OIS spread during both the Financial and European Debt Crisis. Thus, we find a slightly larger impact from the credit component on the LIBOR-OIS spread using our data sample and method. We note that the market for term repo lending, especially the six-month repo loan, is fairly illiquid and thus the observed repo rate may reflect an added market liquidity premium.\footnote{Volume data for term repos published by the OFR can be found at \url{https://www.financialresearch.gov/short-term-funding-monitor/datasets/repo/}} The term repo may therefore contain both market and funding-liquidity risk.\footnote{see \cite{brunnermeier2009market} for details on the close relationship between market and funding-liquidity} Such a premium would result in the model decomposition overestimating the roll-over risk funding-liquidity component in the LIBOR-OIS spread and thus underestimating the size of the credit component.

The figure further plots a regression based decomposition of the credit component. We regress the three- and six-month LIBOR-OIS spread on the equivalent LIBOR-repo spread
\begin{equation}
    Spread^{OIS}_t=\alpha+\beta\times Spread^{repo}_t+\varepsilon_t.
\end{equation}
The estimate of the three-month regression coefficient $\beta^{3m}$ is $0.9652$ $(0.0236)$ and for the six-month spot LIBOR decomposition we obtain $\beta^{6m}$ is $0.8528$ $(0.0314)$. We compute the regression based credit component as $\beta\times Spread^{repo}_t$ to ensure that the credit component spread is zero when the LIBOR-repo spread is zero. The regression shows a similar decomposition to the model mainly diverging in the second half of 2019 where the LIBOR-repo spread went slightly negative. A negative LIBOR-repo spread is inconsistent with our roll-over risk interpretation of the spread and likely due to market liquidity in the term repo market. Our model specification clearly precludes negative LIBOR-repo (as well negative as LIBOR-OIS) spreads, which is also apparent from the model based decomposition of the LIBOR-OIS spread.

\subsection{Comparing Risk Premia in Futures Markets}\label{risk_prem}
Assuming that the LIBOR transition follows its current trajectory and three-month USD LIBOR fixings cease after June 30, 2023 the Eurodollar futures market referencing the LIBOR fixings will also end. Furthermore, the fallback method proposed by CME for all Eurodollar futures contracts maturing after the LIBOR cessation date is to convert Eurodollar futures contracts into the corresponding SOFR futures contracts plus the fixed ISDA fallback spread.\footnote{The CME proposed fallback for Eurodollar contracts is available at \url{https://www.cmegroup.com/content/dam/cmegroup/notices/ser/2021/02/SER-8720.pdf} and \url{https://www.cmegroup.com/education/files/webinar-fallbacks-for-eurodollars.pdf?itm_source=rates_recap_article&itm_medium=hypertext&itm_campaign=rates_recap&itm_content=022021}}
The ISDA fallback spread has already been fixed at $26.161$ basis points for three-month USD LIBOR.\footnote{The fixed fallback spreads can be found at \url{https://assets.bbhub.io/professional/sites/10/IBOR-Fallbacks-LIBOR-Cessation_Announcement_20210305.pdf}}
In this section we compare the risk premia in the new futures market for three-month SOFR contracts to the Eurodollar contracts that these are meant to replace. Any results involving the physical measure are of course impacted by a level of uncertainty due to the fairly large standard deviations of the market price of risk related parameters as displayed in table \ref{table:estimates}.
In order to compare the risk premia in these markets, we consider the zero cost portfolio of buying one Eurodollar futures contract and simultaneously selling the equivalent three-month SOFR futures contract. 
The constructed portfolio should be insensitive to changes in the underlying risk-free rate, but sensitive to the spread processes between SOFR and LIBOR fixings. As previously mentioned, due to the forward-looking nature of the LIBOR rate, the Eurodollar contract settles at the beginning of the interest rate period, while the equivalent SOFR contract settles at the end of the reference period.
The SOFR contract therefore remains risky through the entire period. To account for the mismatch, we calculate the risk premium at time $t$ of the constructed portfolio as the difference between the time $t$ value of the portfolio and the expected value at the beginning of the reference period of the contracts denoted by time $S$.
\begin{equation} \label{rp}
    RP(t;S,T)=f^{ED}(t;S,T)-f^{s,3m}(t;S,T)-E^P\left[f^{ED}(S;S,T)-f^{s,3m}(S;S,T)|\mathcal{F}_t\right]
\end{equation}
As also mentioned in \cite{piazzesi2008futures} Equation \eqref{rp} is not a completely accurate representation of the expected excess return of the portfolio. This is due to the daily mark to market of futures contracts. However, they find that the simplification is insignificant to the size of the risk premium and thus we follow the same approach.

Due to the varying expiry of the actual futures contracts, we cannot compare the risk premium of the portfolio equally across observation dates. Instead, we consider a set of standardized contracts with fixed maturities for each observations date. In particular we will consider contracts for each quarter ahead such that $S-t\in \{90/360,180/360,270/360,360/360\}$.

We note that the completely affine market price of risk specification in Equation \eqref{mpr} results in the risk-neutral mean reversion matrix being identical to the physical mean reversion for the Gaussian part of the state variables. Since we model the SOFR and fed funds futures rates purely as Gaussian processes this implies that the risk premium on these contracts is essentially constant. The time variation in risk premia observed for our constructed portfolio is therefore purely driven by risk premia related to the roll-over risk specific processes. 

\begin{figure}[hbt!]
\centering
\includegraphics[angle=0, scale=.8]{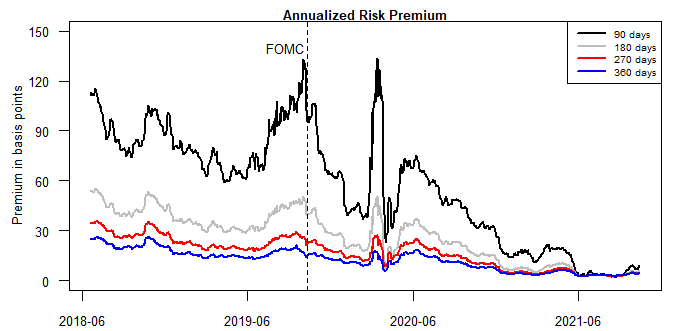}
\caption{The figure plots the annualized risk premium for the constructed portfolio given different holding periods. The vertical line marks the October 11th FOMC meeting.}
\label{fig:risk_premia}
\end{figure}

Figure \ref{fig:risk_premia} plots the resulting annualized risk premium from the strategy.\footnote{Annualized risk premia are obtained by dividing the premium with the holding period of the constructed portfolio, i.e. $S-t$} The portfolio clearly reflects a positive risk premium due to the added risk in the three-month LIBOR fixing compared to the equivalent compounded SOFR rate. the most recent period characterized by the fairly low LIBOR-OIS spread (and equally low LIBOR-SOFR spread) is also reflected in a modest risk premium. The plot also shows that part of the large spike in March 2020 was driven by a significant spike in the risk premia related to interbank lending. Finally, we note that the FOMC announcement on October 11th, was effective in part by successfully reducing the risk premia in interbank lending.

Table \ref{tab:risk_prem} plots the full sample average annualized risk premia for each contract from holding the contract until the beginning of the accrual period. The average annualized risk premium is positive, but decreasing in time to maturity for all contracts. For the one-months contracts we also note that the SOFR and federal funds futures are near identical in risk premium for our sample which is also to be expected given the close to zero $\mu^{\zeta}=-0.1095$ estimate. This aligns with the model-free analysis in \cite{skov2021dynamic} showing very similar excess average returns of one-month SOFR and EFFR futures contracts.

\begin{table}[hbt!]
\centering
\begin{tabular}{ccccc}
$(S-t)$ & 3M SOFR & Eurodollar & 1M SOFR & Federal Funds \\ \hline
90 days & 32.1          & 88.4          & 34.9          & 35.5     \\
180 days & 28.6          & 54.6          & 31.0          & 31.6     \\
270 days & 25.8          & 42.5          & 27.8          & 28.3     \\
360 days & 23.4          & 35.7          & 25.1          & 25.6   \\\hline
\end{tabular}
\caption{Model implied annualized expected excess returns for SOFR, Eurodollar and federal funds futures contracts.}
\label{tab:risk_prem}
\end{table}

\section{Conclusion}
In this paper, we have presented an affine multicurve framework, which jointly models SOFR, EFFR, LIBOR, and term repos while ascribing economic interpretation to the spreads between each of the different rates. The framework thus provides a unified approach for risk managing the LIBOR transition. In particular the model would have many potential use cases such as pricing bespoke tenors, calculating the value transfer of shifting from LIBOR to SOFR as well as pricing nonlinear derivatives involving multiple benchmark rates. 
The LIBOR benchmark allows financial institutions to gain easy exposure to the roll-over risk premium embedded in the cost of funding at term. Such exposure is not easily obtained elsewhere. As we show some of it is obtainable through exposure to credit risky instruments pricing, but a significant component is not, as shown through our decomposition of the LIBOR-OIS spread. The decomposition shows that the spike in the LIBOR-OIS spread during March 2020 was largely driven by an increase in the credit premium, however, during the full sample period we find that the funding-liquidity component is on average equally important in explaining the spread. 
Beyond 2023, as LIBOR is discontinued, the model components would still be identifiable with the LIBOR specific data swapped by one of the likely contenders to replace it. Furthest along is possibly the BSBY, which already has futures contracts traded at the CME, but as of writing with very sparse liquidity. However, even in the absence of a liquid derivatives market there is considerable information gained in simply including spot rate date in the model, as is also demonstrated in this paper. 

\appendix

\section{Pricing in the Affine Setup}
\label{appendix:pricing}
We write the dynamics of the state process under the $Q$-measure on matrix form
\begin{equation}
    dX_t(u)=K^Q \left(\theta^Q-X_t(u) \right)du+\Sigma D(X_t(u),t) dW^Q(u)+dJ_t(u)
\end{equation}
where $X_t(u)=(r^s(u),\theta^s(u),\zeta(u),\lambda_t(u),\phi_t(u),\xi(u),\eta(u),\nu(u))'$ is the vector of state variables.
The drift specifications are given by
\begin{equation} \label{qdrift}
K^Q=\begin{bmatrix}
\kappa^r & -\kappa^r & 0 & 0 & 0 & 0 & 0 & 0 \\
0 & \kappa^\theta & 0 & 0 & 0 & 0 & 0 & 0 \\
0 & 0 & \kappa^\zeta & 0 & 0 & 0 & 0 & 0 \\
0 & 0 & 0 & \beta^\lambda & 0 & 0 & 0 & 0 \\
0 & 0 & 0 & 0 & \beta^\phi & 0 & 0 & 0 \\
0 & 0 & 0 & 0 & 0 & \kappa^\xi & -\kappa^\xi & 0 \\
0 & 0 & 0 & 0 & 0 & 0 & \kappa^\eta & 0 \\
0 & 0 & 0 & 0 & 0 & 0 & 0 & \kappa^\nu 
\end{bmatrix}, \quad
\theta^Q=\begin{bmatrix}
\theta^\theta \\
\theta^\theta \\
\theta^\zeta \\
0 \\
0 \\
\theta^\eta \\
\theta^\eta \\
\theta^\nu
\end{bmatrix}.  
\end{equation}
And the volatility specification becomes
\begin{equation} \label{qvol}
\Sigma=\begin{bmatrix}
\sigma^r & 0 & 0 & 0 & 0 & 0 & 0 & 0 \\
\sigma^\theta \rho & \sigma^\theta\sqrt{1-\rho^2} & 0 & 0 & 0 & 0 & 0 & 0 \\
0 & 0 & \sigma^\zeta & 0 & 0 & 0 & 0 & 0 \\
0 & 0 & 0 & 0 & 0 & 0 & 0 & 0 \\
0 & 0 & 0 & 0 & 0 & 0 & 0 & 0\\
0 & 0 & 0 & 0 & 0 & \sigma^\xi & 0 & 0 \\
0 & 0 & 0 & 0 & 0 & 0 & \sigma^\eta & 0 \\
0 & 0 & 0 & 0 & 0 & 0 & 0 & \sigma^\nu 
\end{bmatrix}, \quad
D(X(t),t)=\text{diag} \left(
\begin{bmatrix}
1 \\
1 \\
1 \\
0 \\
0 \\
\sqrt{\xi(t)} \\
\sqrt{\eta(t)}\\ 
\sqrt{\nu(t)} 
\end{bmatrix}  
\right)
\end{equation}
where $\text{diag}(v)$ is a diagonal matrix with the elements in $v$ on the diagonal. Finally,
$J_t(u)\in \mathbb{R}^8$ and $W^Q(u)\in \mathbb{R}^8$ are defined as
\begin{align}
  &W^Q(u)  =\left(W^r(u),W^\theta(u) ,W^\zeta(u),0,0,W^\xi(u),W^\eta(u),W^\nu(u)\right)',\\
  &J_t(u)  =\left(0,0,0,J_t^\lambda(u),J_t^\phi(u),0,0,0\right)'.
\end{align}
Based on this the ODEs referenced in section \ref{model_theory} and used when computing the futures and spot rates can be expressed in a general framework of ODEs satisfying
\begin{align} \label{ode1start}
\frac{\partial B_1(\tau)}{\partial\tau}&=-\kappa^r B_1(\tau)-\left[R\right]_1, \\
\frac{\partial B_2(\tau)}{\partial\tau}&=-\kappa^\theta B_2(\tau)+\kappa^r B_1(\tau)-\left[R\right]_2,\\
\frac{\partial B_3(\tau)}{\partial\tau}&=-\kappa^\zeta B_3(\tau)-\left[R\right]_3,\\
\frac{\partial B_4(\tau)}{\partial\tau}&=-\beta^\lambda B_4(\tau)-\left[R\right]_4,\\
\frac{\partial B_5(\tau)}{\partial\tau}&=-\beta^\phi B_5(\tau)-\left[R\right]_5,\\
\frac{\partial B_6(\tau)}{\partial\tau}&=-\kappa^\xi B_6(\tau)
+\frac{1}{2}(B_6(\tau)\sigma^\xi)^2
+ \frac{B_4(\tau)}{0.02^{-1}-B_4(\tau)}
-\left[R\right]_6,\\
\frac{\partial B_7(\tau)}{\partial\tau}&=-\kappa^\eta B_7(\tau)+\kappa^\eta B_6(\tau)
+\frac{1}{2}(B_7(\tau)\sigma^\eta)^2-\left[R\right]_7,\\
\frac{\partial B_8(\tau)}{\partial\tau}&=-\kappa^\nu B_8(\tau)
+\frac{1}{2}(B_8(\tau)\sigma^\nu)^2
+ \frac{B_5(\tau)}{0.02^{-1}-B_5(\tau)}
-\left[R\right]_8,\\
\frac{\partial A(\tau)}{\partial\tau}&=(K^Q \theta^Q)' B(\tau)+\frac{1}{2}B(\tau)' \sigma_0\sigma_0' B(\tau),
\label{ode1end}\end{align}
Where $R\in \mathbb{R}^8$ is specific to each of the ODEs and $[R]_i$ denotes the $i^{th}$ element in $R$. $\sigma_0$ is the volatility related to the Gaussian processes
\begin{equation}
    \sigma_0 =
\begin{bmatrix}
P_{3\times3} & 0_{3\times5} \\
0_{5\times3} & 0_{5\times5} 
\end{bmatrix}, \quad 
P_{3\times3}=
\begin{bmatrix}
\sigma^r & 0 & 0 \\
\sigma^\theta \rho & \sigma^\theta\sqrt{1-\rho^2}  & 0\\
0 & 0 & \sigma^\zeta 
\end{bmatrix}.
\end{equation}
\subsection{Spot and Futures Rates}
We now list the different specifications of $R$ used to solve the ODEs presented in section \ref{model_theory}. In order to solve the LIBOR specific ODEs, we note that
\begin{equation}
    E^Q\left[e^{\int_t^T \phi_t(u) du}|\mathcal{F}_t \right]
    =e^{A^{U}(T-t)+B^{U}(T-t)'X_t(t)}
\end{equation}
where $A^{U}(T-t)$ and $B^{U}(T-t)$ solve the general set of ODEs with $R=(0,0,0,0,-1,0,0,0)'$ and initial conditions $A^{U}(0)=B^{U}(0)=0$. Similarly, we calculate the denominator as
\begin{equation}
    E^Q\left[e^{-\int_t^T r^s(u)+\zeta(u)+\lambda_t(u) du}|\mathcal{F}_t \right]
    =e^{A^{Q}(T-t)+B^{Q}(T-t)'X_t(t)}
\end{equation}
and again $A^{Q}(T-t)$ and $B^{Q}(T-t)$ solve the general set of ODEs with $R=(1,0,1,1,0,0,0,0)'$ and initial conditions $A^{Q}(0)=B^{Q}(0)=0$. Finally, the LIBOR specific expressions are calculated as $A^{L}(T-t)=A^{U}(T-t)-A^{Q}(T-t)$ and $B^{L}(T-t)=B^{U}(T-t)-B^{Q}(T-t)$.
Likewise, for the term repo contract we first note that
\begin{equation}
    E^Q\left[e^{-\int_t^T r^s(u) du}|\mathcal{F}_t \right]
    =e^{A^{s}(T-t)+B^{s}(T-t)'X_t(t)}
\end{equation}
where $A^{s}(T-t)$ and $B^{s}(T-t)$ solve the general set of ODEs with $R=(1,0,0,0,0,0,0,0)'$ and initial conditions $A^{s}(0)=B^{s}(0)=0$. Thus, we compute the term repo using the expressions $A^{repo}(T-t)=A^{U}(T-t)-A^{s}(T-t)$ and $B^{repo}(T-t)=B^{U}(T-t)-B^{s}(T-t)$.

The SOFR accumulating account used to price the three-month SOFR futures contract is calculated with $A^{s,3m}(\tau)$ and $B^{s,3m}(\tau)$, which solve the general set of ODEs with $R=(-1,0,0,0,0,0,0,0)'$. The Eurodollar and three-month SOFR futures specific ODEs $A^{ED}(\tau)$ and $B^{ED}(\tau)$ as well as $A^{s,f}(\tau)$ and $B^{s,f}(\tau)$ both solve the general ODEs with $R=(0,0,0,0,0,0,0,0)'$, however, with the different initial conditions given in section \ref{edfut} and \ref{3msofrfut}, respectively. 

For the one-month SOFR futures rate, we first apply the Fubini theorem to change the order of integration
\begin{align}
    \mathbb{E}^Q\left[
    \frac{1}{T-S}\int_S^T r^s(u) du|\mathcal{F}_t\right]
    =
    \frac{1}{T-S}\int_S^T\mathbb{E}^Q\left[ r^s(u)|\mathcal{F}_t\right] du.
\end{align}
We note that the risk-neutral expectation of the Gaussian subset of the state variables is given by
\begin{align}
    \mathbb{E}^Q\left[X_t(u)|\mathcal{F}_t\right]=
\begin{bmatrix}
\theta^\theta +e^{-(u-t)\kappa^r}\left(r(u)-\theta^\theta\right)-\frac{\kappa^r\left(e^{-(u-t)\kappa^r}-e^{-(u-t)\kappa^\theta}\right)}{\kappa^r-\kappa^\theta}\left(\theta(t)-\theta^\theta\right) \\
\theta^\theta +e^{-(u-t)\kappa^\theta}\left(\theta(u)-\theta^\theta\right) \\
\theta^\zeta +e^{-(u-t)\kappa^\zeta}\left(\zeta(u)-\theta^\theta\right) 
\end{bmatrix}.
\end{align}
For the SOFR futures rate, we only need the integral of the first coordinate, which we define as
\begin{align}
    I^r(t;S,T)=&\int_S^T \theta^\theta +e^{-(u-t)\kappa^r}\left(r(u)-\theta^\theta\right)-\frac{\kappa^r\left(e^{-(u-t)\kappa^r}-e^{-(u-t)\kappa^\theta}\right)}{\kappa^r-\kappa^\theta}\left(\theta(t)-\theta^\theta\right) du \nonumber \\
    =&(T-S)\theta^\theta +
\frac{e^{-(S-t)\kappa^r}-e^{-(T-t)\kappa^r}}{\kappa^r}
\left(r(u)-\theta^\theta\right)
\nonumber\\&+\frac{\kappa^r\left(
e^{-(S-t)\kappa^\theta}-e^{-(T-t)\kappa^\theta}\right)
-\kappa^\theta\left(
e^{-(S-t)\kappa^r}-e^{-(T-t)\kappa^r}\right)}{\kappa^\theta(\kappa^r-\kappa^\theta)}
\left(\theta(u)-\theta^\theta\right).
\end{align}
Thus, the expectation is
\begin{align}
    \mathbb{E}^Q\left[
    \frac{1}{T-S}\int_S^T r^s(u) du|\mathcal{F}_t\right]=\frac{1}{T-S} I^r(t;S,T)
\end{align}
For the federal funds futures rate, we include the integral over the expectation of the SOFR-EFFR spread process
\begin{align}
    I^\zeta(t;S,T)&=\int_S^T \theta^\zeta +e^{-(u-t)\kappa^\zeta}\left(\zeta(u)-\theta^\theta\right)du \nonumber \\ 
    &=(T-S)\theta^\zeta +
\frac{e^{-(S-t)\kappa^\zeta}-e^{-(T-t)\kappa^\zeta}}{\kappa^\zeta}
\left(\zeta(u)-\theta^\zeta\right).
\end{align}
Such that the expectation is calculated as
\begin{align}
    \mathbb{E}^Q\left[
    \frac{1}{T-S}\int_S^T r^{FF}(u) du|\mathcal{F}_t\right]=\frac{1}{T-S} \left(I^r(t;S,T)+I^\zeta(t;S,T)\right).
\end{align}

\subsection{Swap Rates}
This section presents the results required to compute the swap rates in section \ref{swap_comp}. On October 16, 2020, the London Clearing House (LCH) and Chicago Mercantile Exchange (CME) changed the Price Alignment Interest (PAI) on cleared swaps from EFFR to SOFR. For the sake of simplicity, we assume SOFR to be the collateral rate for the entire sample period.\footnote{See \cite{rutkowski2021pricing} for an in depth study on pricing SOFR derivatives under differential funding cost.} To ease the notation, we define the pseudo zero coupon bond referencing SOFR $p^s(t,T):=e^{-\int_t^T r^s(u)du}$, which we compute as
\begin{equation}
    p^s(t,T)=e^{A^{s}(T-t)+B^{s}(T-t)'X_t(t)},
\end{equation}
where we reiterate that $A^{s}(T-t)$ and $B^{s}(T-t)$ solve the Riccati equations with $R=(1,0,0,0,0,0,0,0)'$ and initial conditions $A^{s}(0)=0$ and $B^{s}(0)=0$.
The floating leg SOFR-rate is calculated from the daily compounded SOFR analogously to the three-month SOFR futures contract in Equation \eqref{sofr3mfut}. Again, we consider the continuous approximation, which we denote $R^{s}(T_{i-1},T_i)$. 
In the USD market SOFR OIS payments occur at a similar frequency for each leg and once a year for all contracts with a maturity of one year or greater. We denote the payment dates by $T_i$ for $i=1,...,n$ and $\delta=T_i-T_{i-1}$ with $T_n=T$. The fair SOFR OIS rate, $OIS^s(t,T)$, is then given by
\begin{align}
    OIS^s(t,T)&=\frac
    {\sum_{i=1}^n \delta\mathbb{E}^Q\left[e^{-\int_t^{T_i} r^s(u)du}R^{s}(T_{i-1},T_i) |\mathcal{F}_t\right]}
    {\sum_{i=1}^n \delta p^s(t,T_i)} \nonumber \\
    &=\frac
    {\sum_{i=1}^n p^s(t,T_{i-1})-p^s(t,T_i)}
    {\sum_{i=1}^n \delta p^s(t,T_i)} \nonumber\\
    &=\frac
    {1-p^s(t,T_n)}
    {\sum_{i=1}^n \delta p^s(t,T_i)} 
\end{align}
Applying the same continuous approximation for the standard OIS contract referencing EFFR, the resulting OIS rate satisfies
\begin{equation}
    OIS^{FF}(t,T)=\frac
    {\sum_{i=1}^n \delta\mathbb{E}^Q\left[e^{-\int_t^{T_i} r^s(u)du}R^{FF}(T_{i-1},T_i) |\mathcal{F}_t\right]}
    {\sum_{i=1}^n \delta p^s(t,T_i)}.
\end{equation}
Where $R^{FF}(T_{i-1},T_i)$ denotes the continuous approximation of the floating leg EFFR-rate.
We note that OIS contracts referencing SOFR and EFFR follow identical conventions.
In order to calculate the expectation in the nominator we rewrite it as
\begin{align}
    &\mathbb{E}^Q\left[e^{-\int_t^{T_i} r^s(u)du}R^{FF}(T_{i-1},T_i) |\mathcal{F}_t\right] \nonumber\\
    &=\mathbb{E}^Q\left[e^{-\int_t^{T_i} r^s(u)du}\frac{1}{\delta}\left(e^{\int_{T_{i-1}}^{T_i} r^{FF}() du}-1\right) |\mathcal{F}_t\right] \nonumber\\
    &=\frac{1}{\delta}\mathbb{E}^Q\left[e^{-\int_t^{T_i} r^s(u)du} e^{\int_{T_{i-1}}^{T_i} r^{FF}(u) du}|\mathcal{F}_t\right] - \frac{p^s(t,T_i)}{\delta}.
\end{align}
Focusing on the term on the left we have
\begin{align}
    &\frac{1}{\delta}\mathbb{E}^Q\left[e^{-\int_t^{T_i} r^s(u)du} e^{\int_{T_{i-1}}^{T_i} r^{FF}(u) du}|\mathcal{F}_t\right] \nonumber\\
    &=\frac{1}{\delta}\mathbb{E}^Q\left[e^{-\int_t^{T_{i-1}} r^s(u)du} e^{\int_{T_{i-1}}^{T_i} r^{FF}(u)-r^{s}(u) du}|\mathcal{F}_t\right] \nonumber\\
    &=\frac{1}{\delta}\mathbb{E}^Q\left[e^{-\int_t^{T_{i-1}} r^s(u)du} \mathbb{E}^Q\left[e^{\int_{T_{i-1}}^{T_i} \zeta(u) du}|\mathcal{F}_{T_{i-1}}\right]
    |\mathcal{F}_t\right].
\end{align}
The inner expectation can be solved using the general set of ODEs where $R=(0,0,-1,0,0,0,0,0)'$
\begin{equation}
    \mathbb{E}^Q\left[e^{\int_{T_{i-1}}^{T_i} \zeta(u) du}|\mathcal{F}_{T_{i-1}}\right]
    =
    e^{A^{aux}(\delta)+B^{aux}(\delta)X_{T_{i-1}}(T_{i-1})}
\end{equation}
with initial conditions $A^{aux}(0)=0$ and $B^{aux}(0)=0$. This allows us to compute the entire expectation as
\begin{align}
    \mathbb{E}^Q\left[e^{-\int_t^{T_{i-1}} r^s(u)du} e^{A^{aux}(\delta)+B^{aux}(\delta)'X_{T_{i-1}}(T_{i-1})}
    |\mathcal{F}_t\right]
    =e^{A^{OIS}(T_{i-1}-t)+B^{OIS}(T_{i-1}-t)X_t(t)}
\end{align}
where $A^{OIS}(T_{i-1}-t)$ and $B^{OIS}(T_{i-1}-t)$ again solve the general set of ODEs with $R=(1,0,0,0,0,0,0,0)$ and initial conditions $A^{OIS}(0)=A^{aux}(\delta)$ and $B^{OIS}(0)=B^{aux}(\delta)$.

Next, we consider an IRS referencing the LIBOR fixings. The standard USD IRS references the three-month LIBOR with the floating leg payment frequency matching the LIBOR fixing. The payment dates on the fixed leg are semi-annually. We denote the floating leg payment dates $T^{L}_i$ for $i=1,...,n$ with $T^{L}_i-T^{L}_{i-1}=\delta^L$ and the fixed leg payment dates $T^{fix}_i$ for $i=1,...,m$ with $T^{fix}_i-T^{fix}_{i-1}=\delta^{fix}$. Finally, we assume identical dates for the final payment such that $T_n=T_M=T$. The IRS rate ensuring zero initial value of the swap is 
\begin{equation}
    IRS(t,T)=\frac
    {\sum_{i=1}^n \delta^L\mathbb{E}^Q\left[e^{-\int_t^{T^{L}_i} r^s(u)du}L(T^{L}_{i-1},T^{L}_i) |\mathcal{F}_t\right]}
    {\sum_{i=1}^m \delta^{fix}p^s(t,T_i)}.
\end{equation}
Following the approach of the OIS we rewrite the expectation as
\begin{align}
    \mathbb{E}^Q\left[e^{-\int_t^S r^s(u)du}L(T^{L}_{i-1},T^{L}_i) |\mathcal{F}_t\right]
    =\frac{1}{\delta^L}\mathbb{E}^Q\left[e^{-\int_t^S r^s(u)du} \frac{U(T^{L}_{i-1},T^{L}_i)}{Q(T^{L}_{i-1},T^{L}_i)}|\mathcal{F}_t\right] - \frac{p^s(t,T^{L}_i)}{\delta^L}.
\end{align}
Focusing on the term on the left, we have
\begin{align}
    &\frac{1}{\delta^L}\mathbb{E}^Q\left[e^{-\int_t^{T^{L}_i} r^s(u)du} \frac{U(T^{L}_{i-1},T^{L}_i)}{Q(T^{L}_{i-1},T^{L}_i)}|\mathcal{F}_t\right] \nonumber\\
    =&\frac{1}{\delta^L}\mathbb{E}^Q\left[\mathbb{E}^Q\left[e^{-\int_t^{T^{L}_i} r^s(u)du} \frac{U(T^{L}_{i-1},T^{L}_i)}{Q(T^{L}_{i-1},T^{L}_i)}
    |\mathcal{F}_{T^{L}_{i-1}}\right]|\mathcal{F}_t\right] \nonumber\\
    =&\frac{1}{\delta^L}\mathbb{E}^Q\left[e^{-\int_t^{T^{L}_{i-1}} r^s(u)du} p^s(T^{L}_{i-1},T^{L}_i)\frac{U(T^{L}_{i-1},T^{L}_i)}{Q(T^{L}_{i-1},T^{L}_i)}
    |\mathcal{F}_t\right].
\end{align}
The expectation is then calculated similarly to the Eurodollar futures rate in Equation \eqref{eurodollar}. Thus, we compute the term as
\begin{align}
    \frac{1}{\delta^L}e^{A^{IRS}(T^{L}_{i-1}-t)+B^{IRS}(T^{L}_{i-1}-t)'X_t(t)},
\end{align}
where $A^{IRS}(T^{L}_{i-1}-t)$ and $B^{IRS}(T^{L}_{i-1}-t)$ solve the general ODEs with $R=(1,0,0,0,0,0,0,0)'$ and initial conditions given by $A^{IRS}(0)=A^{s}(\delta^L)+A^{U}(\delta^L)-A^{Q}(\delta^L)$ and $B^{IRS}(0)=B^{s}(\delta^L)+B^{U}(\delta^L)-B^{Q}(\delta^L)$ except for the the jump elements with $J_t^\lambda(u)$ and $J_t^\phi(u)$, which have their initial conditions set equal to zero reflecting the renewal of the LIBOR panel.
\subsubsection{Credit Default Swaps}
We calculate the fair CDS spread on a unit notional contract providing protection over the period $[t;T]$ while assuming zero recovery at default. The value of the protection leg is
\begin{equation}
    \mathbb{E}\left[e^{-\int_t^{\tau_t} r^s(u)du} 1_{(\tau_t\le T)} |\mathcal{G}_t\right],
\end{equation}
and for the payment leg
\begin{align} \label{cds_sum1}
    &\sum_{i=1}^n
    C(T_i-T_{i-1})
    \mathbb{E}\left[e^{-\int_t^{T_i} r^s(u)du} 1_{(T_i<\tau_t)} |\mathcal{G}_t\right]  \\\label{cds_sum2}
    &+
    \sum_{i=1}^n
    C(\tau_t-T_{i-1})
    \mathbb{E}\left[e^{-\int_t^{\tau_t} r^s(u)du} 1_{(T_{i-1}<\tau_t \le T_i)} |\mathcal{G}_t\right],
\end{align}
where $C$ denotes the spread, $T_0=t$ and $T_i$ for $i=1,..,n$ are the quarterly payment dates with $T_n=T$. 
Following the approach in \cite{backwell2019term} we calculate the value of the protection leg as
\begin{align}
    \mathbb{E}\left[e^{-\int_t^{\tau_t} r^s(u)du} 1_{(\tau_t\le T)} |\mathcal{G}_t\right]
    &=\lim_{m\to\infty}\sum_{i=1}^m
    \mathbb{E}\left[e^{-\int_t^{\tau_t} r^s(u)du} 1_{(t_{i-1} <\tau_t\le t_i)} |\mathcal{G}_t\right] \nonumber\\
    &=\lim_{m\to\infty}\sum_{i=1}^m
    \mathbb{E}\left[e^{-\int_t^{t_{i-1}} r^s(u)du} 1_{(t_{i-1} <\tau_t\le t_i)} |\mathcal{G}_t\right]
\end{align}
Applying the identity $1_{(t_{i-1} <\tau_t\le t_i)}=1_{(t_{i-1} <\tau_t)}-1_{(t_{i} <\tau_t)}$ and the intensity based credit risk approach from Appendix \ref{appendix:intensity} Equation \eqref{intensity} we have
\begin{align}
&\lim_{m\to\infty}\sum_{i=1}^m
    \mathbb{E}\left[e^{-\int_t^{t_{i-1}} r^s(u)du} \left(e^{-\int_t^{t_{i-1}} \Lambda(u)+\lambda_t(u)du}
    -e^{-\int_t^{t_{i}}\Lambda(u)+ \lambda_t(u)du}\right) |\mathcal{F}_t\right]
    \nonumber\\
&=\lim_{m\to\infty}\sum_{i=1}^m
    \mathbb{E}\left[e^{-\int_t^{t_{i-1}} r^s(u)du} 
    (\Lambda(u)+\lambda_t(u))
    e^{-\int_t^{t_{i}} \Lambda(u)+\lambda_t(u)du} (t_i-t_{i-1})|\mathcal{F}_t\right]    
\end{align}
Where we have used the relation for the increments $f(t_{i-1})-f(t_{i})=-f'(t_{i-1})(t_i-t_{i-1})$ with $f(x)=e^{-\int_t^{x} \Lambda(u)+\lambda_t(u)du}$. Rewriting as an integral the value of the protection leg can be calculated as
\begin{align}
\int_t^T
    \mathbb{E}\left[\lambda_t(u) e^{-\int_t^u r^s+\lambda_t(u)du}
     |\mathcal{F}_t\right]du.
\end{align}
When calculating the integrand we note that $\Lambda(u)$ is unidentified in our model and thus omitted in the calculation of the CDS spread (see section \ref{swap_comp}), that is we set $\Lambda(u)=0$. The integrand is the computed using the extended transform in \cite{duffie2000transform}  
\begin{equation} \label{integrand}
    \left(a(u-t)+b(u-t)'X_t(t)\right)e^{A^{cds}(u-t)+B^{cds}(u-t)'X_t(t)}
\end{equation}
where $A^{cds}(u-t)$ and $B^{cds}(u-t)$ solve the ODEs in Equation \eqref{ode1start})-\eqref{ode1end} with initial conditions $A^{cds}(0)=0$ $B^{cds}(0)=0$ and $R=(1,0,0,1,0,0,0,0)'$ while $a(\tau)$ and $b(\tau)$ satisfy the modified ODEs
\begin{align}
\frac{\partial b_1(\tau)}{\partial\tau}&=-\kappa^r b_1(\tau), \\
\frac{\partial b_2(\tau)}{\partial\tau}&=-\kappa^\theta b_2(\tau)+\kappa^r b_1(\tau),\\
\frac{\partial b_3(\tau)}{\partial\tau}&=-\kappa^\zeta b_3(\tau),\\
\frac{\partial b_4(\tau)}{\partial\tau}&=-\beta^\lambda b_4(\tau),\\
\frac{\partial b_5(\tau)}{\partial\tau}&=-\beta^\phi b_5(\tau),\\
\frac{\partial b_6(\tau)}{\partial\tau}&=-\kappa^\xi b_6(\tau)
+\frac{1}{2}B_6(\tau)b_6(\tau)(\sigma^\xi)^2
+ \frac{0.02^{-1} b_4(\tau)}{(0.02^{-1}-B_4(\tau))^2},\\
\frac{\partial b_7(\tau)}{\partial\tau}&=-\kappa^\eta b_7(\tau)+\kappa^\eta B_6(\tau)
+\frac{1}{2}B_7(\tau)b_7(\tau)(\sigma^\eta)^2,\\
\frac{\partial b_8(\tau)}{\partial\tau}&=-\kappa^\nu b_8(\tau)
+\frac{1}{2}B_8(\tau)b_8(\tau)(\sigma^\nu)^2
+ \frac{0.02^{-1} b_5(\tau)}{(0.02^{-1}-B_5(\tau))^2},\\
\frac{\partial a(\tau)}{\partial\tau}&=(K^Q \theta^Q)' b(\tau)+\frac{1}{2}B(\tau)' \sigma_0\sigma_0' b(\tau),
\end{align}
with initial conditions $a(0)=0$ and $b(0)=(0,0,0,1,0,0,0,0)'$. We can then numerically integrate the integrand in \eqref{integrand} to obtain the value of the protection leg.
For the payment leg we calculate the expectation in the first sum in Equation \eqref{cds_sum1} as
\begin{equation}
     \mathbb{E}\left[e^{-\int_t^{T_i} r^s(u)du} 1_{(T_i<\tau_t)} |\mathcal{G}_t\right]=
     e^{A^{cds}(T_i-t)+B^{cds}(T_i-t)'X_t(t)}.
\end{equation}
For the expectation in the second sum \eqref{cds_sum2} we apply the same approach as in the protection leg
\begin{align}
    &\mathbb{E}\left[e^{-\int_t^{\tau_t} r^s(u)du} 1_{(T_{i-1}<\tau_t \le T_i)} |\mathcal{G}_t\right]
    \nonumber \\
    &=\int_{T_{i-1}}^{T_{i}}
    (u-T_{i-1})
        \left(a(u-t)+b(u-t)'X_t(t)\right)e^{A^{cds}(u-t)+B^{cds}(u-t)'X_t(t)} du.
\end{align}

\section{The Intensity Based Credit Risk Approach}\label{appendix:intensity}
In order to price credit risky instruments, we follow \cite{filipovic2013term} and extend the doubly stochastic framework to accommodate default times for each arbitrary time $t$. Thus, we consider the filtration $\mathcal{F}_t=\sigma (X(s) | 0 \leq s \leq t )$ and an i.i.d. sequence of exponentially distributed random variables $E_t\sim \text{Exp}(1)$ independent of the filtration $\mathcal{F}_t$. 
For each $t$ we then define the random default times by
\begin{align}
\tau_t=\text{inf}\left\{T>t|\int_{t}^T \Lambda(u)+\lambda_t (u) du \geq E_t \right\},
\end{align}
and denote the filtration generated by all default indicator processes as $\mathcal{H}_t$. The random times $\tau_t$ are stopping times with respects to the enlarged filtration $\mathcal{G}_t=\mathcal{F}_t \vee \mathcal{H}_t$. Following standard results on the intensity based approach (see e.g. \cite{lando2009credit}) and assuming zero recovery, any
$\mathcal{F}_T$-measurable integrable random variable, $X$, satisfies
\begin{align} \label{intensity}
\mathbb{E}^Q\left[X 1_{(\tau_t>T)}|\mathcal{G}_t\right]
=\mathbb{E}^Q\left[X e^{-\int_t^T \Lambda(u)+\lambda_t(u) du}|\mathcal{F}_t\right].
\end{align}

\section{The Term Repo Approximation}\label{appendix:repo}
In this appendix we investigate the accuracy of the approximation in Equation \eqref{repo_term_approx} to the true term repo rate, which we recall is given by
\begin{equation}
    R^{repo}(t,T)=\frac{1}{T-t}\left( \frac{E^Q\left[e^{\int_t^T\psi(u)+\phi_t(u)du}|\mathcal{F}_t \right]}{E^Q\left[e^{-\int_t^T r(u) du}|\mathcal{F}_t \right]}-1\right).
\end{equation}
It is clear from this expression that we are unable compute the repo rate without identifying the underlying risk-free rate, $r(u)$ and the repo specific risk premium, $\psi(u)$, separately. However, applying Jensen's inequality, we can construct both an upper and lower bound for $R^{repo}(t,T)$ using only processes identified in our model, which allows us to evaluate the maximum possible error of the approximation in \eqref{repo_term_approx}.

First, we note that the approximation used in the paper is in fact a lower bound. To see this, assume independence between $\psi(u)$ and $\phi_t(u)$ as well as $r(u)$ and $\psi(u)$ as in section \ref{section:TermRepo}. Focusing on the fraction we have
\begin{align} \label{lowerbound}
    \frac{E^Q\left[e^{\int_t^T\psi(u)+\phi_t(u)du}|\mathcal{F}_t \right]}{E^Q\left[e^{-\int_t^T r(u) du}|\mathcal{F}_t \right]}
    &=
    \frac{
    E^Q\left[e^{\int_t^T\phi_t(u)du}|\mathcal{F}_t \right]
    E^Q\left[e^{\int_t^T\psi(u)du}|\mathcal{F}_t \right]}
    {E^Q\left[e^{-\int_t^T r(u) du}|\mathcal{F}_t \right]}
    \nonumber
    \\&=
    \frac{E^Q\left[e^{\int_t^T\phi_t(u)du}|\mathcal{F}_t \right]}
    {E^Q\left[e^{-\int_t^T r(u) du}|\mathcal{F}_t \right]
    \frac{1}{
    E^Q\left[e^{\int_t^T\psi(u)du}|\mathcal{F}_t \right]}}
    \nonumber
    \\&\geq
    \frac{E^Q\left[e^{\int_t^T\phi_t(u)du}|\mathcal{F}_t \right]}
    {E^Q\left[e^{-\int_t^T r(u) du}|\mathcal{F}_t \right]
    E^Q\left[\frac{1}{
    e^{\int_t^T\psi(u)du}}|\mathcal{F}_t \right]}
    \nonumber
    \\&=
    \frac{E^Q\left[e^{\int_t^T\phi_t(u)du}|\mathcal{F}_t \right]}
    {E^Q\left[e^{-\int_t^T r^s(u) du}|\mathcal{F}_t \right]}.
\end{align}
Next, in order to calculate an upper bound, we assume independence between $\psi(u)+\phi_t(u)$ and $r(u)$ which yields
\begin{align}\label{upperbound}
    \frac{E^Q\left[e^{\int_t^T\psi(u)+\phi_t(u)du}|\mathcal{F}_t \right]}{E^Q\left[e^{-\int_t^T r(u) du}|\mathcal{F}_t \right]}
    &=
    E^Q\left[e^{\int_t^T\psi(u)+\phi_t(u)du}|\mathcal{F}_t \right]
    \frac{1}{E^Q\left[e^{-\int_t^T r(u) du}|\mathcal{F}_t \right]} \nonumber
    \\ &\leq
    E^Q\left[e^{\int_t^T\psi(u)+\phi_t(u)du}|\mathcal{F}_t \right]
    E^Q\left[\frac{1}{e^{-\int_t^T r(u) du}}|\mathcal{F}_t \right] \nonumber
    \\ &=
    E^Q\left[e^{\int_t^T r^s(u)+\phi_t(u)du}|\mathcal{F}_t \right].
\end{align}
Thus, combining Equation \eqref{lowerbound} and \eqref{upperbound} we obtain
\begin{equation}
    \frac{E^Q\left[e^{\int_t^T\phi_t(u)du}|\mathcal{F}_t \right]}
    {E^Q\left[e^{-\int_t^T r^s(u) du}|\mathcal{F}_t \right]}
    \leq
    \frac{E^Q\left[e^{\int_t^T\psi(u)+\phi_t(u)du}|\mathcal{F}_t \right]}{E^Q\left[e^{-\int_t^T r(u) du}|\mathcal{F}_t \right]}
    \leq
    E^Q\left[e^{\int_t^T r^s(u)+\phi_t(u)du}|\mathcal{F}_t \right]
\end{equation}
Because of the Gaussian specification of the $r^s(u)$ and $\zeta(u)$ processes, the spread between the upper and lower bound is constant across the entire sample for each of the considered maturities. Specifically, given our model estimates, we find that the spread between the upper and lower bound approximation of the term repo rate is $0.02$ basis points for the three-month repo and $0.08$ basis points for the six-month repo.
Furthermore, due to the low volatility of the $\zeta(u)$ process (see table \ref{table:estimates}), we expect the underlying $\psi(u)$ to display an equally low volatility. This implies that the ignored convexity adjustment in \eqref{lowerbound} is negligible and thus the lower bound is a very close approximation to the actual term repo rate.

\section{The Kalman Filter}\label{appendix:kalman}
We estimate the model using quasi-maximum likelihood method with the Kalman Filter on spot LIBOR and end of day futures data.
When estimating the time series property of the model, we consider the reduced state process $X(t)\in \mathbb{R}^6$ excluding the roll-over risk jump variables and thus we omit the subscript. Having defined a completely affine market price of risk the implied drift parameters under the $P$-measure are obtained from 
\begin{align}
    &K^P=\hat{K}^Q-\hat{\Sigma} \text{diag}(\mu)\delta^2, \\
    &\theta^P=(K^P)^{-1}[\hat{K}^Q\hat{\theta}^Q+\hat{\Sigma} \text{diag}(\mu)\delta^1]
\end{align}
where $\hat{\theta}^Q$, $\hat{K}^Q$ and $\hat{\Sigma}$ refer to the $Q$-parameters in Equation \eqref{qdrift} and \eqref{qvol} excluding the fourth and fifth row and column specific to the roll-over risk jump variables and
\begin{align}
\mu=
\begin{bmatrix}
\mu^r \\
\mu^\theta \\
\mu^\zeta \\
\mu^\xi \\
\mu^\eta \\
\mu^\nu
\end{bmatrix}, \quad
\delta^1=
\begin{bmatrix}
1 \\
1 \\
1 \\
0\\
0 \\
0 
\end{bmatrix}, \quad
\delta^2 =\text{diag} \left(
\begin{bmatrix}
0 \\
0 \\
0 \\
1\\
1 \\
1 
\end{bmatrix}  
\right).
\end{align}
We now consider the discretized state process under the $P$-dynamics given by
\begin{align}
X(t_i)=\theta^P+e^{-K^P \Delta t} (X(t_{i-1})-\theta^P) + \int_{t_{i-1}}^{t_{i}}  e^{-K^P u}\hat{\Sigma} D(X(u),u) dW^{P}(u)
\end{align}
with $\Delta t=t_{i}-t_{i-1}$ set to $1/252$ to approximately reflect the number of daily futures data observations in a year.
Rearranging we define the state equation of the Kalman Filter as
\begin{align}
X(t_{i})=\left( I-e^{-K^P \Delta t}\right)\theta^P + e^{-K^P \Delta t} X(t_{i-1}) + \omega(t_{i})
\end{align}
with $\omega(t_{i}) \sim \mathcal{N}(0,Z(t_{i}))$. The conditional covariance matrix, $Z(t_{i})$, is equal to
\begin{align}
Z(t_{i})=\int_{t_{i-1}}^{t_{i}}  e^{-K^P u}\hat{\Sigma} D(\mathbb{E}\left[X(u)|\mathcal{F}(t_{i-1})\right],u) 
D(\mathbb{E}\left[X(u)|\mathcal{F}(t_{i-1})\right],u)' 
\hat{\Sigma}' \left(e^{-K^P u}\right)'du.
\end{align}
We compute $Z(t_{i})$ analytically using the methods presented in \cite{analyticalsecond}.
The prediction step of the kalman filter is then 
\begin{align}
&X(t_{i}|t_{i-1})=F X(t_{i-1}|t_{i-1}) + C, \\
&P(t_{i}|t_{i-1})=F P(t_{i-1}|t_{i-1}) F' + Z(t_{i}).
\end{align}
Where $F=e^{-K^P \Delta t} $ denotes the state transition model and $C=\left( I-e^{-K^P \Delta t}\right)\theta^P$ the control input. The standard Kalman filter requires the measurement equation to be affine in the state vector
\begin{align}
y(t_{i}) = A(t_{i})+B(t_{i})' X(t_{i}|t_{i-1})+\varepsilon(t_{i})
\end{align}
with $\varepsilon(t_{i}) \sim \mathcal{N}(0,H)$ where $H$ is a diagonal matrix, and the transition and measurement errors, $\omega(t_{i})$ and $\varepsilon(t_{i})$, are independent.
The measurement residuals are computed using the time $t$ futures and spot LIBOR yields data, $y(t_{i})$, and the model implied yields
\begin{align}
\nu_t = y(t_{i})-\left( A(t_{i})+B(t_{i})' X(t_{i}|t_{i-1})\right).
\end{align}
The residuals have a conditional mean of $0$ and conditional variance given by
\begin{align}
S(t_{i}):=\mathbb{V}[\nu(t_{i})|y(t_{i-1}),...,y(t_{1})]=H+B(t_{i}) P(t_{i}|t_{i-1}) B(t_{i})'.
\end{align}
In the update step the a priori state estimates are updated using the observed time $t_{i}$ data
\begin{align}
&X(t_{i}|t_{i})=X(t_{i}|t_{i-1}) +K(t_{i}) \nu(t_{i}), \\
&P(t_{i}|t_{i})=(I-K(t_{i}) B(t_{i})) P(t_{i}|t_{i-1}).
\end{align}
$K(t_{i})$ is the optimal Kalman gain matrix
\begin{align}
K(t_{i}) = P(t_{i}|t_{i-1}) B(t_{i}) ' (S(t_{i}))^{-1}.
\end{align}
Assuming a Gaussian quasi log-likelihood for a given set of parameters, $\Theta$, the log-likelihood is determined by the conditional mean and variance of the innovations $\nu(t_{i})$
\begin{align}
l(y(t_{1}),...,y(t_{T});\Theta)=\sum_{i=1}^T \left(- \frac{N}{2}\log(2\pi) - \frac{1}{2} \left(\log |S(t_{i})| +\nu(t_{i}) ' (S(t_{i}))^{-1} \nu(t_{i}) \right)\right)
\end{align}
with $N$ the number of observations at each date and $T$ the amount of dates in out data sample.
To obtain the optimal set of parameters, $\hat{\Theta}$, we maximize the log-likelihood function using the Nelder-Mead algorithm with a function value tolerance of $0.01$. The state variables are required to be stationary under the $P$-measure and we therefore require the eigenvalues of $K^P$ to be positive. \\
The futures market calendar differs slightly from the LIBOR publishing dates. In the estimation we consider all dates with futures resulting in dates with missing spot LIBOR data. Furthermore, the six-month repo rate is rather illiquid and thus there are days in our time series for which we have no six-month repo quote, particularly around March 2020. The missing observations are easily accommodated in the kalman filter by considering an $N(t_{i}) \times N$-matrix $W(t_{i})$ where $N$ is the total amount of observations including missings and $N(t_{i})$ is the actual amount of observations at time $t_{i}$. $W(t_{i})$ is then a subset of $I_N$ such that $y^\star(t_{i})=W(t_{i}) y(t_{i})$ is the set of non-missing observations. The measurement equation is then simply replaced by
\begin{align}
y^\star(t_{i}) = A^\star(t_{i})+B^\star(t_{i})' X(t_{i}|t_{i-1})+\varepsilon^\star(t_{i}), \quad \varepsilon^\star(t_{i}) \sim \mathcal{N}(0,H^\star(t_{i})),
\end{align}
where $A^\star(t_{i})=W(t_{i})A(t_{i})$, $B^\star(t_{i})=W(t_{i})B(t_{i})$ and $H^\star(t_{i})=W(t_{i}) H W(t_{i})'$.
Finally, we change $N$ in the likelihood contributions to reflect the actual number of observations on a given date $N(t_{i})$.

\bibliographystyle{chicago}
\spacingset{1}
\bibliography{Refs}

\end{document}